\documentclass[aps,amsmath,amssymb,showpacs,showkeys]{revtex4}
\usepackage[dvips]{graphicx,color}

\usepackage[utf8]{inputenc}
\usepackage{multirow}
\usepackage{amsmath}
\usepackage{times}
\usepackage{xcolor}
\usepackage{mathtools}
\usepackage[%
  colorlinks=true,
  urlcolor=blue,
  linkcolor=red,
  citecolor=blue
]{hyperref}

\newcommand{\orcid}[1]{\href{https://orcid.org/#1}{\includegraphics[width=8pt]{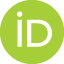}}}

\begin{document}
\title{Quasinormal modes and thermodynamic properties of GUP-corrected 
Schwarzschild black hole surrounded by quintessence}

\author{Ronit Karmakar \orcid{0000-0002-9531-7435}}
\email[Email: ]{ronit.karmakar622@gmail.com}

\affiliation{Department of Physics, Dibrugarh University,
Dibrugarh 786004, Assam, India}
\author{Dhruba Jyoti Gogoi \orcid{0000-0002-4776-8506}}
\email[Email: ]{moloydhruba@yahoo.in}

\affiliation{Department of Physics, Dibrugarh University,
Dibrugarh 786004, Assam, India}

\author{Umananda Dev Goswami  \orcid{0000-0003-0012-7549}}
\email[Email: ]{umananda2@gmail.com}

\affiliation{Department of Physics, Dibrugarh University,
Dibrugarh 786004, Assam, India}

\begin{abstract}

We study the Quasinormal Modes (QNMs) of the Schwarzschild black hole 
surrounded by a quintessence field after implementing the quantum corrections 
to its solution as required by the Generalised Uncertainty Principle (GUP). We 
analyse the dependence of the QNMs on the deformation parameters of GUP as 
well as on the quintessence parameter. In most cases the QNMs show the 
appreciable dependency on these parameters. For a better idea of the accuracy 
of calculations of QNMs, we compare the 
results of the QNMs obtained via Mashhoon method with the 6th order WKB 
method. A good agreement between these two methods of QNM calculations is seen 
depending on the different factors. Further, we study the thermodynamic 
properties of the GUP-corrected Schwarzschild black hole and check for any 
dependence with the deformation parameters and the quintessence parameter. 
In particular, we compute the Hawking temperature, heat capacity and entropy 
for the black hole and analyse the results graphically to show the dependency 
of the thermodynamic properties on the said parameters. We have seen that 
the thermodynamic properties of black holes also depend noticeably on the model
parameters in most cases. Black hole remnants have been studied and it is 
shown that the possible existence of remnant radius as well as remnant 
temperature depends on the deformations introduced. However, it is observed 
that the GUP-corrected black hole constructed here can not become a remnant. 


\end{abstract}


\maketitle
\section{Introduction}
Black holes and Gravitational Waves (GWs) are two of the most fascinating 
predictions of General Relativity (GR). Black holes are among the most 
mysterious objects in our universe that have attracted the attention of 
scientific community for many decades. It is widely believed that when a 
sufficiently massive star runs out of fuel, the inward pull of gravity becomes 
dominant and there is a rapid collapse of matter towards the centre, which 
ultimately results in the formation of a black hole. A black hole interacts 
with its surrounding matter and is generally in a perturbed state. Such 
perturbations cause the black hole to undergo oscillation and that leads to 
the emission of GWs \cite{ligo_1,konoplya_1, konoplya_2, michele_1, 
hughes}. GWs are ripples in the spacetime fabric, generated by accelerating 
massive objects, which propagate at the speed of light. The LIGO-Virgo 
collaboration declared the first-ever detection of GWs on 14th of september 
2015 \cite{abbott2016}, almost after 100 years of their prediction by prominent physicist A.\ 
Einstein in 1916. Since then, a number of detections of GWs has been reported 
by the collaboration \cite{abbott2016_2,abbott2017, abbott2020,abbott2021}. It 
has generated a new era of gravitational astronomy, the likes of which has 
never been seen before.

Quasinormal Modes (QNMs) are some complex frequencies associated with the GWs 
that represent the reaction of a black hole, after some perturbations act on it 
\cite{mashhoon_1, chandrasekhar_1, konoplya_1, konoplya_2, konoplya_3, 
konoplya_4, djgogoi_1, djgogoi_2, djgogoi_3}. There are different methods of
calculations of QNMs from the black holes \cite{konoplya_1}. One of the 
simplest and elegant methods of finding out the QNMs was devised by B.\ 
Mashhoon, which is commonly called as the Mashhoon Method \cite{konoplya_1,
 mashhoon_1,mashhoon_2,mashhoon_3, mashhoon_4}. It is an analytical method 
which is easy to handle for a simple system. The most frequently used and 
trusted method of QNM analysis is the Wentzel-Kramers-Brillouin (WKB) method 
\cite{konoplya_1,konoplya_2,  iyer}, which was initially utilized by Schutz 
and Will \cite{schutz}. Improvements were made in this method by Konoplya who 
introduced corrections in WKB calculations upto 6th orders \cite{konoplya_4}. 
Recently, more higher order corrections are introduced in this method 
(\cite{konoplya_5} and references therein). In this work we consider both these methods of analyzing 
the QNMs for the sake of accuracy in the calculation. Apart from these, 
numerous works have been done on the analytical and numerical techniques to 
calculate the QNMs associated with the black hole perturbations in recent 
times \cite{leaver_1, leaver_2, cardoso, karlos,zhidenko, zhidenko2,ali1,ali2}.
Other works related to black hole physics may be found in \cite{nandan,kala} 
and in the references therein.

Recently, the black hole physics and related thermodynamics have also 
attracted researcher's attention and a lot of such works can be seen in 
recent times in literature \cite{marco,p,ted,zhang,michael,marco_1,M,ch,yuan,
sharif_1, mohsen,faizal1}. It was after the ground-breaking works of Bekenstein and Hawking that cemented the idea of interpreting the black hole as a 
thermodynamic system, showing the corresponding properties like temperature 
and entropy. Hawking proposed that a black hole emits radiation from its event 
horizon \cite{hawking_1,hawking_2, hawking_3} and Bekenstein showed 
that as the black hole engulfs matter, the information associated with it is 
not lost but is incorporated in the horizon area of the black hole 
\cite{bekenstein_1,bekenstein_2, bekenstein_3}. These ideas led to a 
revolution in black hole thermodynamics.  

The discovery of Riess and Perlmutter, which showed that the universe is 
expanding with an acceleration \cite{riess,perlmutter} led to a flood of 
theoretical models coming forward to resolve this unexpected observation. 
In the majority of such models the idea of dark energy was invoked to interpret 
this colossal expansion (see \cite{dark} for a review). One of the convenient 
ideas to deal with it was the $\Lambda$CDM model, in which Einstein's 
cosmological constant was reintroduced as a homogeneous and isotropic fluid 
with the negative pressure and is considered to be the cause of this present 
state of expansion of the universe \cite{dark_2}. The cosmological constant 
was thought to originate out of quantum fluctuations of vacuum, but its 
theoretically predicted value could not match the observational value. To 
alleviate this problem, dynamical scalar field models were proposed, and 
the most common among  them is the quintessence model of dark energy. In this
model a scalar field minimally coupled to gravity is used to describe this 
late time accelerated expansion. Detailed about and current status of 
the quintessence model can be seen in Refs.~\cite{dark,dark_2,q1,q2,q3,q4,q5}. 

Many works can be found in literature where the black hole thermodynamics 
is studied with a surrounding field. In 2003, a Schwarzschild black hole 
surrounded by a quintessence was studied and corresponding thermodynamic 
properties were examined by Kiselev \cite{kiselev}. He showed that presence 
of the surrounding field has a major impact on the properties of a black hole. 
Subsequently, Chen \textit{et al.}~\cite{chen_1} considered a d-dimensional 
black hole with a quintessence matter surrounding it and examined 
thermodynamic properties of the black hole. Reissner-Nordstr\"om black holes 
were examined with a quintessential surrounding by Wei and Chu \cite{wei_1}. 
Thermodynamics of Narai type black holes were considered by Fernando with 
a quintessential surrounding \cite{fernando}. Recently, it is seen that many 
research works have considered the effects of quantum corrections via the 
Generalised Uncertainty Principle (GUP) on the thermodynamics of black hole 
\cite{shahjalal,anacleto_1,bcl}. One such work was carried out by Shahjalal 
in 2019 \cite{shahjalal}, where he compared the effects of the quantum 
deformations with and without the presence of quintessential surroundings. The 
case of rotating non-linear magnetically charged black holes was taken up by 
Ndongmo \textit{et al.}~\cite{ndongmo}, in which thermodynamics of the black 
hole was studied. Moreover, Anacleto \textit{et al.} \cite{anacleto_1} studied 
the quantum-corrected Schwarzschild black holes and analysed the absorption 
and scattering processes. Gonz\'ales \textit{et al.} \cite{gonzales} studied 
a 3-dimensional Godel black hole and calculated the QNMs and Hawking radiation.
Further, L\"utf\"uoglu \textit{et al.}~\cite{bcl} studied the thermodynamics of 
Schwarzschild black holes with the quintessential surrounding and GUP. They 
showed that the upper and lower bounds on various functions like temperature 
and entropy depend on the deformation parameters as well as on the quintessence 
coefficient, and also presented plots of P-V isotherms. It is notable that the 
combined linear and quadratic GUP approach was introduced in \cite{n1,n2,n3,n4,n5,n6}.

The study of black hole shadows has also been carried out in recent times as 
they provide useful insights into the black hole event horizon as well as into 
the optical properties of a black hole \cite{konoplya_shadow,vagnozzi}. The 
initial work in this direction was done by Synge \cite{synge} and Luminet
\cite{luminet} in the  1970s and for rotating Kerr black holes by Bardeen 
\cite{bardeen1}. Lately, an interesting work on QNMs and shadows of 
Schwarzschild black hole was performed by Anacleto 
\textit{et al.}~\cite{anacleto}, where 
they considered the GUP-modified Schwarzschild black hole solutions and 
calculated the QNMs and shadows of the black holes, and showed the dependency 
of these properties on the deformation parameters. In recent times, many works 
have been done in this field \cite{j1,s1,s2,s3,s4,s5,s7,s8,s9,s10,s11,s12,s13}. 
In this work, however, we shall not include the shadow analysis and 
it will be addressed in a future work.

Thus, inspired from the ongoing endeavours to explore these novel ideas, in 
this work we intend to study the various properties of a GUP-corrected 
Schwarzschild black hole surrounded by a quintessence field, such as the QNMs 
and thermodynamic properties like Hawking temperature, entropy, heat capacity 
and surface gravity. To the best of our information, GUP with both linear and 
quadratic terms has not been incorporated to Schwarzschild black hole 
surrounded by quintessence. The novelty of GUP that it introduces a minimum 
length scale, that is the Planck's length, might play an important role in the 
properties of black holes \cite{bcl,anacleto,faizal2,faizal3}. 

The rest of the paper is organized as follows. In section II, we compute the 
QNMs of the GUP-corrected Schwarzschild black hole surrounded by a quintessence
field using the Mashhoon method and the 6th order WKB method, and analyse
 the results. In section III, we study the thermodynamic properties of the 
black hole and present a graphical analysis of dependency of the thermodynamic 
properties on the deformation parameters as well as on the quintessence 
coefficient. We summarize our results and present some concluding 
remarks in section IV.

\section{Quasinormal modes of a GUP-corrected Schwarzschild black hole}
The general form of the black hole metric as initially derived by Kiselev 
\cite{kiselev}, in which he considered a Schwarzschild black hole surrounded by 
a quintessence dark energy with a particular energy density, can be expressed 
by
\begin{equation}
ds^2 = -g(r)\, dt^2 + \frac{1}{g(r)}\, dr^2 + r^2\, d\Omega^2,
\label{eq1}
\end{equation}
where $d\Omega^2 = d\theta^2+\sin^2\theta\, d\phi^2$ and the metric function 
$g(r)$ has the from:
\begin{equation}
g(r) = 1 - \frac{2M}{r}-\frac{e}{r^{3\omega +1}}.
\label{eq2}
\end{equation}
In this function $M$ is the mass of the black hole, $\omega$ is the equation 
of state parameter of the quintessence field, and $e$ is the positive 
normalization coefficient that is dependent on the quintessence density. 
Since the recent past, a quantum correction to various black hole solutions,
including the Schwarzschild one has been introduced via GUP in order to avoid 
singularities in such solutions by introducing a minimum length other than zero \cite{kaz}. 
Under this correction, the normal metric of a black hole is modified, which gives a corresponding 
new horizon of the black hole. Thus, for example, in the case of 
Schwarzschild black hole, the original Schwarzschild horizon radius $r_{h}$ of the black hole 
has to be replaced with the GUP-corrected radius $r_{hGUP}$ for this 
purpose \cite{anacleto}. The basic steps of incorporation of GUP correction 
into the black hole metric \eqref{eq1} are the following. 

\begin{figure}[!h]
\includegraphics[scale=0.26]{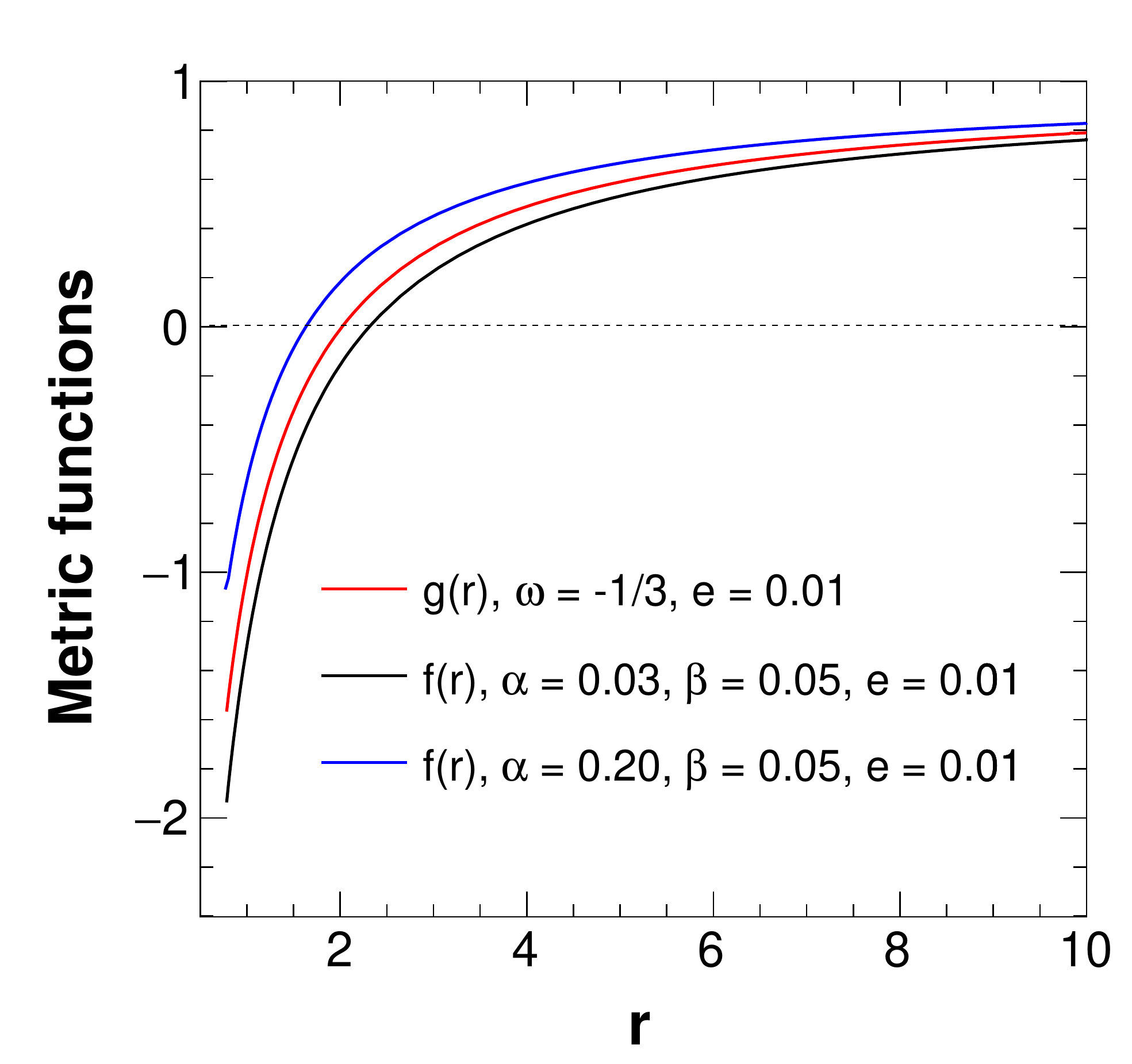}\hspace{0.3cm}
\includegraphics[scale=0.26]{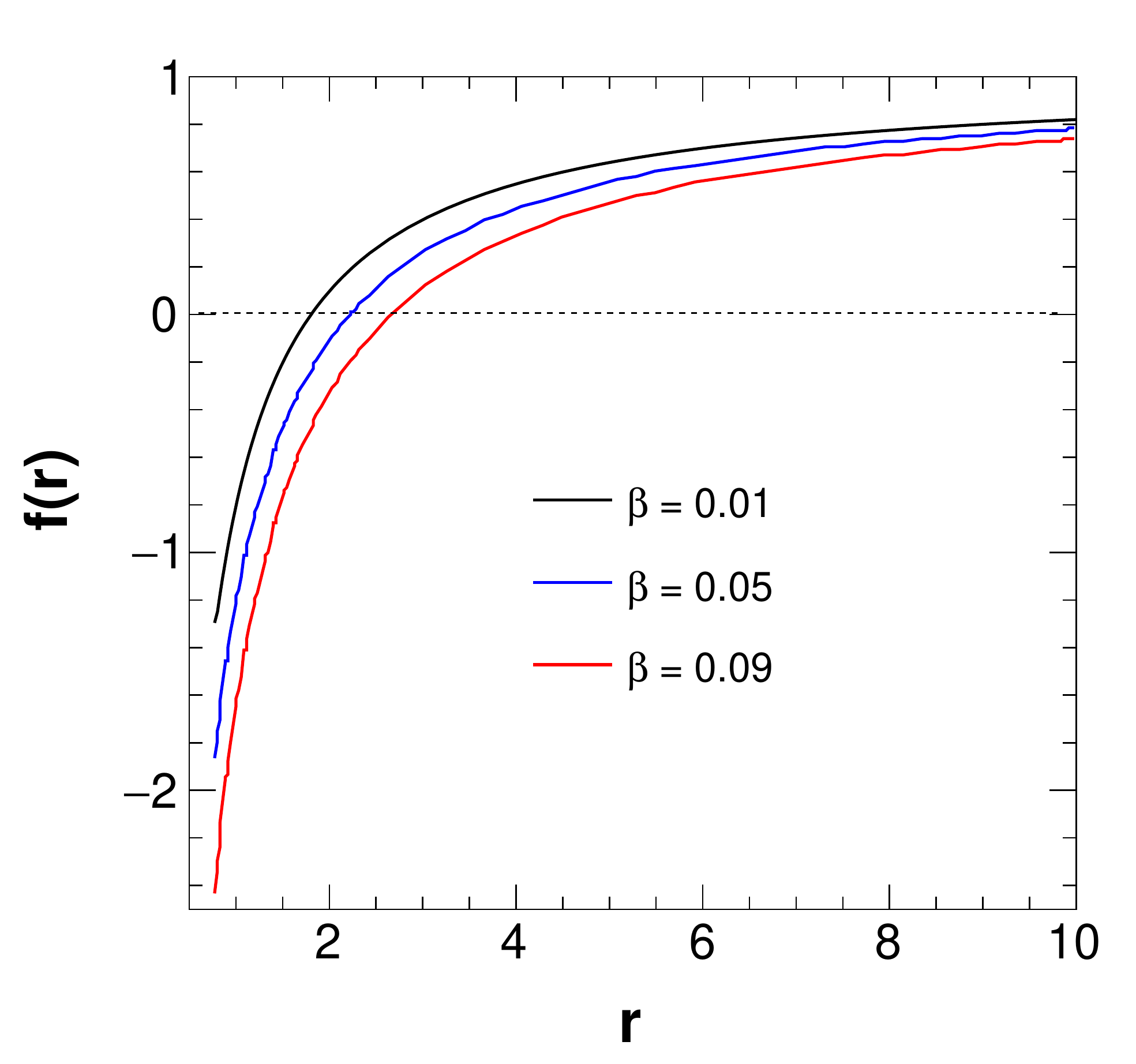}\hspace{0.3cm}
\includegraphics[scale=0.26]{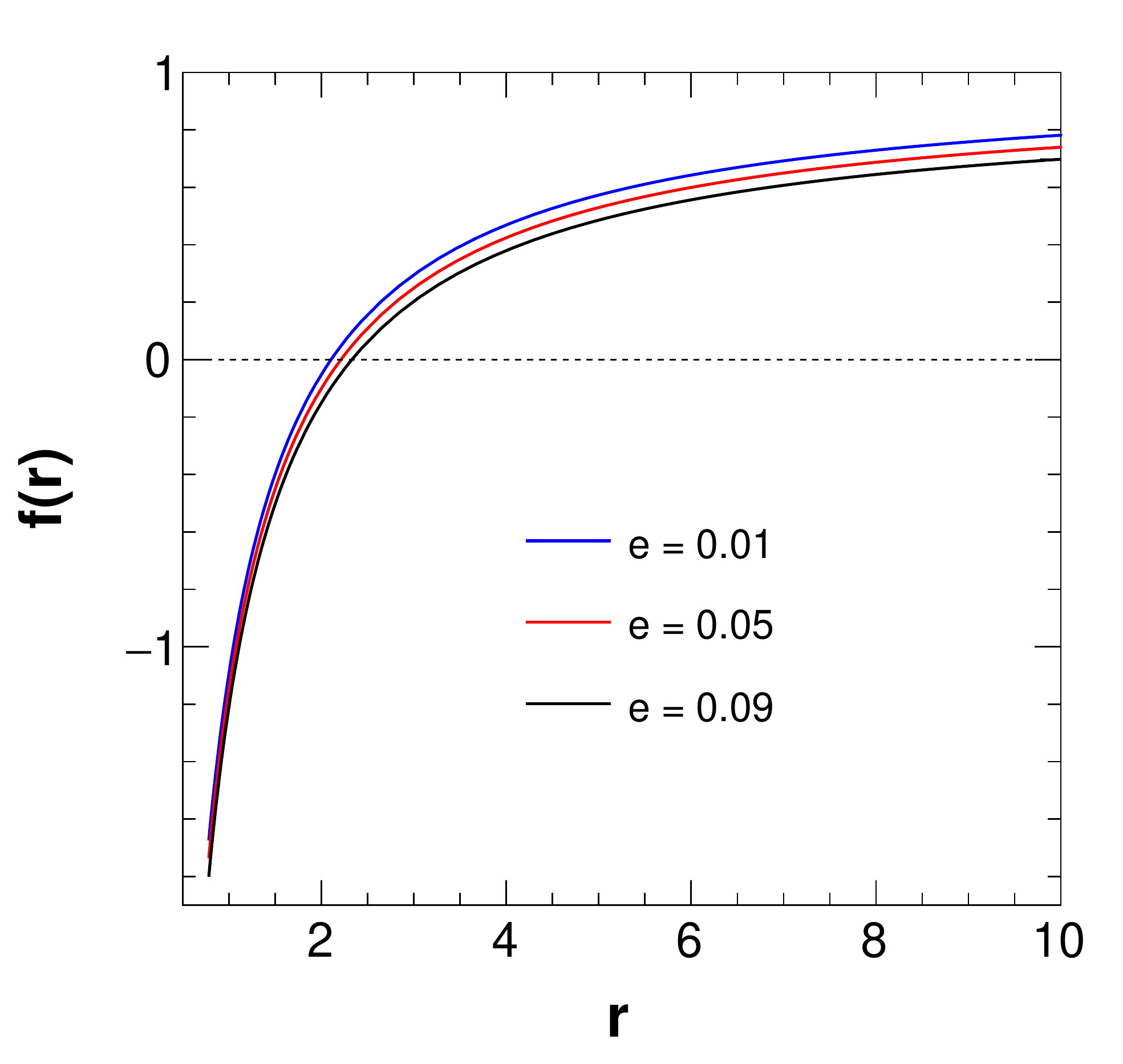}
\caption{Behaviours of the original metric and the GUP-modified metric as a 
function of $r$ for different values of the associated parameters as shown.
In the middle plot $\alpha=0.09$ is used, whereas in the right plot 
$\alpha = \beta = 0.02$ is used. In all the three plots $M=1$ is used. This 
value of $M$ is used for all plots and analysis of this work, if otherwise not 
stated.}
\label{fig1}
\end{figure}

Considering the modified Heisenburg algebra, we may write
\begin{equation}
\Delta x \Delta p \geq \frac{\hbar}{2}\bigg(1-\frac{\alpha l_p}{\hbar} \Delta p + \frac{\beta l^2_p}{\hbar^2} (\Delta p)^2 \bigg),
\label{eq3}
\end{equation}
where $\alpha$ and $\beta$ are two dimensionless deformation parameters, and 
$l_p$ is the Planck's length. Taking the unit system with $G=c=\hbar=l_p =1$, 
equation \eqref{eq3} can be solved for $\Delta p$, which gives
\begin{equation}
\Delta p \approx \frac{4r_h+\alpha}{2\beta}\left[1-\sqrt{1-\frac{4\beta}{(4r_h+\alpha)^2}}\right].
\label{eq4}
\end{equation}
Here the uncertainty in position is taken as the horizon diameter $2r_h$ and we 
end up with the following expression \cite{anacleto_1}:
\begin{equation}
E_{GUP} \geq E\left[1-\frac{4\alpha}{r_h}+\frac{16 \beta}{r_h^2}+\dots \right],
\label{eq5}
\end{equation}
where $E_{GUP}$ is the GUP-corrected energy of the black hole. Now considering 
the assumption that $E\sim M$, $E_{GUP}\sim M_{GUP}$ and calculating 
$r_h =\frac{2M}{1-e}$ from the function \eqref{eq2}, we obtained the relation,
\begin{equation}
M_{GUP}\geq M\left[1-\frac{4\alpha}{r_h}+\frac{16 \beta}{r_h^2}\right]=M\left(1-\frac{2\alpha(1-e)}{M}+\frac{4 \beta(1-e)^2}{M^2}\right).
\label{eq6}
\end{equation}
Thus, finally the line element of a GUP-corrected Schwarzschild black hole 
surrounded by a quintessence field ($\omega = -1/3$) takes the form:
\begin{equation}
ds^2 = -f(r)dt^2 + \frac{1}{f(r)}dr^2 + r^2 d\Omega^2
\label{eq7}
\end{equation}
with the modified metric function 
\begin{equation}
f(r)=1-\frac{2M}{r}\left(1-\frac{2\alpha(1-e)}{M}+\frac{4 \beta(1-e)^2}{M^2}\right)-e \equiv 1 -\frac{2M_{GUP}}{r} - e.
\label{eq8}
\end{equation}
This metric function gives the GUP-corrected horizon radius as $r_{hGUP} = 
2M_{GUP}/(1-e)$. Fig.~\ref{fig1} shows the behaviours of the original 
metric and the GUP-corrected metric as functions of $r$ for various values of 
the related parameters. From the figure it is seen that there is only one 
event horizon for the black hole for different values of the parameters. 
There is no other horizon obtained for the black hole. The left plot shows 
that with the increasing values of $\alpha$, the horizon radius becomes 
smaller. Whereas the middle and the right plots show that with the increasing 
$\beta$ and $e$ values respectively, the horizon radius increases. It 
is also seen that the effect of parameters $\alpha$ and $\beta$ is identical 
and dominant than that of the parameter $e$. 

At this stage it is necessary to mention that the small values of the GUP 
parameters is an obvious choice because any correction term that we introduce 
into our theory cannot be larger than the base term involved as these 
corrections are generally very minute in nature. In the literature, we found 
that the values of these parameters have been considered less than unity 
\cite{bcl,anacleto,shahjalal,gogoi4}. This choice is well motivated as we are considering only 
small corrections to the original uncertainty relation and it is demanded for 
the derivation of the metric expression. Similarly, the quintessence 
parameter ($e$) has been constrained for the case of a Schwarzschild black 
hole surrounded by a quintessence field in Ref.~\cite{A6}, where the authors found a 
bound on the quintessential parameter as $10^{-21}\leq eM\leq 10^{-11}$. On 
the other hand in Refs.\ \cite{chen_1,toledo}, it can be seen that the 
quintessential parameter is taken of the order $\sim 0.1$. So, in our study
we consider reasonably small values of these parameters.

\subsection{QNMs by Mashhoon Method}
Mashhoon method is an analytical method of calculating the QNMs of a black 
hole by comparing its effective potential with a standard potential, such as 
the Poschl-Teller potential \cite{mashhoon_2} or the Eckart potential 
\cite{mashhoon_1}. In this work we consider the scalar field 
perturbation for the calculation of QNMs from the black hole of our model. In
this perturbation method the equation of motion for a test scalar field $\Phi$, which becomes perturbed according to the perturbation of the black hole 
spacetime, is considered and has the form:
\begin{equation}
\frac{1}{\sqrt{g}} \partial_{p}(\sqrt{-g} g^{pq}\partial_{q})\Phi^p_q=0.
\label{eq9}
\end{equation}
It is appropriate to express the scalar field $\Phi$ as spherical 
harmonic waves in the form:
\begin{equation}
\Phi^p_q(t,r,\theta,\phi)=e^{-i \omega t} \frac{\psi(r)}{r}Y^p_q(\theta,\phi),
\label{eq10}
\end{equation}
where $\psi(r)$ represents the radial part, $Y^p_q$ represents the 
spherical harmonics and $\omega$ represents the oscillation frequency for 
the time part of the wave. $\omega$ is actually the quasinormal frequency 
which we want to find out using two different methods. Using equation 
\eqref{eq10} in \eqref{eq9}, we can conveniently transform the equation of 
motion into a Schr\"odinger-like wave equation given by
\begin{equation}
\frac{d^2 \psi}{dx^2}+\big(\omega^2-V_l\big)\psi=0.
\label{eq11}
\end{equation}
Here, the tortoise coordinate $x$ is defined as $dx=dr/f(r)$ and $V_l(r)$ 
is the effective black hole potential, which can be 
obtained from the formula \cite{djgogoi_2,konoplya_5}
\begin{equation}
V_l(r)=f(r) \Big[\frac{f'(r)}{r}+\frac{l(l+1)}{r^2}\Big].
\label{eq12}
\end{equation}
After obtaining the effective potential of the black hole, we compare it with 
the standard Poschl-Teller potential at their maxima. The Poschl-Teller 
potential has the form \cite{mashhoon_2}:
\begin{equation}
V_{PT}=\frac{V_{0}}{\cosh^2 a(x-x_{0})},
\label{eq13}
\end{equation} 
where the quantity $V_{0}$ denotes the height and $a$ denotes the curvature 
of the potential at its maximum. On comparing the two potentials, we get the 
analytical form of the parameters $V_0$ and $a$ (see the Appendix for
the explicit form of their expressions). Then we utilize the formulae 
for calculating the QNMs, which is given by Mashhoon as \cite{mashhoon_2}
\begin{equation}
\omega=\omega_{R} + i \omega_{I}=\pm \Big(V_0 - \frac{a^2}{4}\Big)^{\frac{1}{2}}+ i a\Big(n+\frac{1}{2}\Big),
\label{eq14}
\end{equation}
where $\omega_{R}$ and $\omega_{I}$ are the real and imaginary parts 
of the QNM frequency respectively.
Now, using this formula (the explicit form of the formula can be found
in the appendix) we have calculated the QNM frequencies of the GUP-corrected 
black hole surrounded by a quintessential dark energy field as shown in the 
Table \ref{table01}. In this calculation of QNMs, we have considered a small 
positive value of the quintessence parameter $e =0.05$ laying well within the
accepted range and mass of the black hole is considered as $M=1$. The 
quantum deformation parameters $\alpha$ and $\beta$ are also assumed as small 
positive values within their well accepted range as shown in the table. 

\begin{table}[h!]
\caption{QNMs of the GUP-corrected black hole surrounded by a 
quintessence field for $n=0$, $n=1$, $n=2$ modes, for multipole number 
$l=1$, quintessence parameter $e=0.05$ and various values of the deformation parameter $\alpha$ and $\beta$ obtained by using the Mashhoon method.}
\vspace{2mm}
\centering
\begin{tabular}{c@{\hskip 5pt}c@{\hskip 10pt}c@{\hskip 10pt}c@{\hskip 10pt}c@{\hskip 10pt}c@{\hskip 10pt}c@{\hskip 10pt}c@{\hskip 5pt}c}
\hline\hline
&$\alpha$ & $\beta$ & $e$ &  QNMs for $n=0$ & QNMs for $n=1$ &  QNMs for $n=2$ & QNMs for $n=3$ & \\
\hline\hline
&0.00 & 0.00 & 0.05 & 0.269699 + 0.106170i & 0.269699 + 0.318511i & 0.269699 + 0.530852i & 0.269699 + 0.743192i& \\
&0.00 & 0.01 & 0.05 & 0.257500 + 0.109324i & 0.257500 + 0.327971i & 0.257500 + 0.546619i & 0.257500 + 0.765267i& \\
&0.00 & 0.02 & 0.05 & 0.246116 + 0.111819i & 0.246116 + 0.335457i & 0.246116 + 0.559095i & 0.246116 + 0.782733i& \\
&0.00 & 0.03 & 0.05 & 0.235482 + 0.113762i & 0.235482 + 0.341287i & 0.235482 + 0.568812i & 0.235482 + 0.796337i& \\ 
\hline 
&0.02 & 0.00 & 0.05 & 0.283504 + 0.101997i & 0.283504 + 0.305990i & 0.283504 + 0.509984i & 0.283504 + 0.713977i& \\
&0.02 & 0.01 & 0.05 & 0.270366 + 0.105984i & 0.270366 + 0.317951i & 0.270366 + 0.529919i & 0.270366 + 0.741886i& \\
&0.02 & 0.02 & 0.05 & 0.258121 + 0.109175i & 0.258121 + 0.327526i & 0.258121 + 0.545876i & 0.258121 + 0.764227i& \\
&0.02 & 0.03 & 0.05 & 0.246696 + 0.111702i & 0.246696 + 0.335107i & 0.246696 + 0.558512i & 0.246696 + 0.781916i& \\
\hline
&0.04 & 0.00 & 0.05 & 0.298387 + 0.096763i & 0.298387 + 0.290290i & 0.298387 + 0.483817i & 0.298387 + 0.677343i& \\
&0.04 & 0.01 & 0.05 & 0.284222 + 0.101762i & 0.284222 + 0.305286i & 0.284222 + 0.508810i & 0.284222 + 0.712334i& \\
&0.04 & 0.02 & 0.05 & 0.271034 + 0.105795i & 0.271034 + 0.317385i & 0.271034 + 0.528975i & 0.271034 + 0.740565i& \\
&0.04 & 0.03 & 0.05 & 0.258744 + 0.109025i & 0.258744 + 0.327075i & 0.258744 + 0.545124i & 0.258744 + 0.763174i& \\
\hline
&0.06 & 0.00 & 0.05 & 0.314447 + 0.090238i & 0.314447 + 0.270713i & 0.314447 + 0.451188i & 0.314447 + 0.631663i& \\
&0.06 & 0.01 & 0.05 & 0.299161 + 0.096470i & 0.299161 + 0.289410i & 0.299161 + 0.482350i & 0.299161 + 0.675289i& \\
&0.06 & 0.02 & 0.05 & 0.284942 + 0.101525i & 0.284942 + 0.304574i & 0.284942 + 0.507623i & 0.284260 + 0.712247i&\\
&0.06 & 0.03 & 0.05 & 0.271705 + 0.105604i & 0.271705 + 0.316812i & 0.271705 + 0.528020i & 0.271705 + 0.739228i&\\
\hline
\end{tabular}
\label{table01}
\end{table}
It is interesting to note that there is a striking dependence of the QNMs on 
the deformation parameters. As is seen from Table \ref{table01}, when $\alpha$ 
is kept constant and $\beta$ is increased, there is a noticeable decrease in 
the magnitude of the real part of the QNMs. That is, the amplitude of QNMs is 
inversely proportional to $\beta$. Whereas, with a particular value of 
$\beta$, it is seen that the amplitude increases with increase in $\alpha$. 
On the other hand in the case of the imaginary part of the QNM representing the 
damping of the wave, for a fixed value of $\alpha$, the damping increases with 
increase in $\beta$. While, for a fixed $\beta$, the damping decreases with an 
increase in $\alpha$. We have also calculated the QNMs for the black hole with 
$l=2$ as shown in Table \ref{table02}. Similar pattern is observed in 
this case also. But one notable feature that comes out is that with an increase 
in $l$, the corresponding amplitudes of the QNMs increase noticeably, while 
the damping factor is not much affected although it decreases with increasing
$l$. Another feature that is apparent from 
all the calculated modes is that with the increase in $n$ values the amplitude
of QNMs remains the same but its damping increases. This is in fact already 
clear from equation \eqref{eq11}.  

\begin{table}[h!]
\caption{QNMs of GUP-corrected black hole surrounded by a
 quintessence field for $n=0$, $n=1$, $n=2$ and $n=3$ modes, for multipole 
number $l=2$, quintessence parameter $e=0.05$ and various values of the 
deformation parameter $\alpha$ and $\beta$ obtained by using the Mashhoon 
method.}
\vspace{2mm}
\centering
\begin{tabular}{c@{\hskip 5pt}c@{\hskip 10pt}c@{\hskip 10pt}c@{\hskip 10pt}c@{\hskip 10pt}c@{\hskip 10pt}c@{\hskip 10pt}c@{\hskip 5pt}c}
\hline \hline
&$\alpha$ & $\beta$  & $e$ & QNMs for $n=0$ & QNMs for $n=1$ & QNMs for $n=2$ & QNMs for $n=3$& \\
\hline \hline
&0.00 & 0.00 & 0.05 & 0.447498 + 0.102677i & 0.447498 + 0.308031i & 0.447498 + 0.513384i & 0.447498 + 0.718738i & \\ 
&0.00 & 0.01 & 0.05 & 0.430431 + 0.105325i & 0.430431 + 0.315974i & 0.428658 + 0.527870i & 0.430431 + 0.737273i &\\
&0.00 & 0.02 & 0.05 & 0.414525 + 0.107389i & 0.414525 + 0.322167i & 0.414525 + 0.536946i & 0.414525 + 0.751724i &\\
&0.00 & 0.03 & 0.05 & 0.399674 + 0.108966i & 0.399674 + 0.326897i & 0.399674 + 0.544828i & 0.399674 + 0.762759i &\\
\hline 
&0.02 & 0.00 & 0.05 & 0.466854 + 0.099131i & 0.466854 + 0.297394i & 0.466854 + 0.495657i & 0.466854 + 0.693919i &\\
&0.02 & 0.01 & 0.05 & 0.448431 + 0.102519i & 0.448431 + 0.307558i & 0.448431 + 0.512596i & 0.448431 + 0.717635i &\\
&0.02 & 0.02 & 0.05 & 0.431299 + 0.105201i & 0.431299 + 0.315602i & 0.431299 + 0.526004i & 0.431299 + 0.736406i &\\
&0.02 & 0.03 & 0.05 & 0.415335 + 0.107293i & 0.415335 + 0.321880i & 0.415335 + 0.536467i & 0.415335 + 0.751054i &\\
\hline 
&0.04 & 0.00 & 0.05 & 0.487794 + 0.094643i & 0.487794 + 0.283930i & 0.487794 + 0.473216i & 0.487794 + 0.662503i &\\
&0.04 & 0.01 & 0.05 & 0.467862 + 0.098931i & 0.467862 + 0.296793i & 0.467862 + 0.494654i & 0.467862 + 0.692516i &\\
&0.04 & 0.02 & 0.05 & 0.449367 + 0.102360i & 0.449367 + 0.307079i & 0.449367 + 0.511798i & 0.449367 + 0.716518i &\\
&0.04 & 0.03 & 0.05 & 0.432170 + 0.105075i & 0.432170 + 0.315226i & 0.432170 + 0.525376i & 0.432170 + 0.735527i &\\
\hline 
&0.06 & 0.00 & 0.05 & 0.510501 + 0.089004i & 0.510501 + 0.267012i & 0.510501 + 0.445020i & 0.510501 + 0.623028i &\\
&0.06 & 0.01 & 0.05 & 0.488885 + 0.094391i & 0.488885 + 0.283172i & 0.488885 + 0.471953i & 0.488885 + 0.660734i &\\
&0.06 & 0.02 & 0.05 & 0.468874 + 0.098728i & 0.468874 + 0.296184i & 0.468874 + 0.493640i & 0.468874 + 0.691096i &\\
&0.06 & 0.03 & 0.05 & 0.450307 + 0.102198i & 0.450307 + 0.306594i & 0.450307 + 0.510991i & 0.450307 + 0.715387i &\\
\hline 
\end{tabular}
\label{table02}
\end{table}

It will also be interesting to observe any dependence of the QNMs with the 
quintessence parameter $e$, which has been kept at a constant value in the 
previous analysis. For this purpose, as shown in the Table \ref{table03}, we 
have computed the QNMs for different values of the parameter $e$ with some 
fixed values of the deformation parameters $\alpha$ and $\beta$. It is seen 
from the table that for a fixed value of $l$, with increasing $e$, the 
amplitude decreases, while the damping increases.

\begin{figure}[h!]
\includegraphics[scale=0.31]{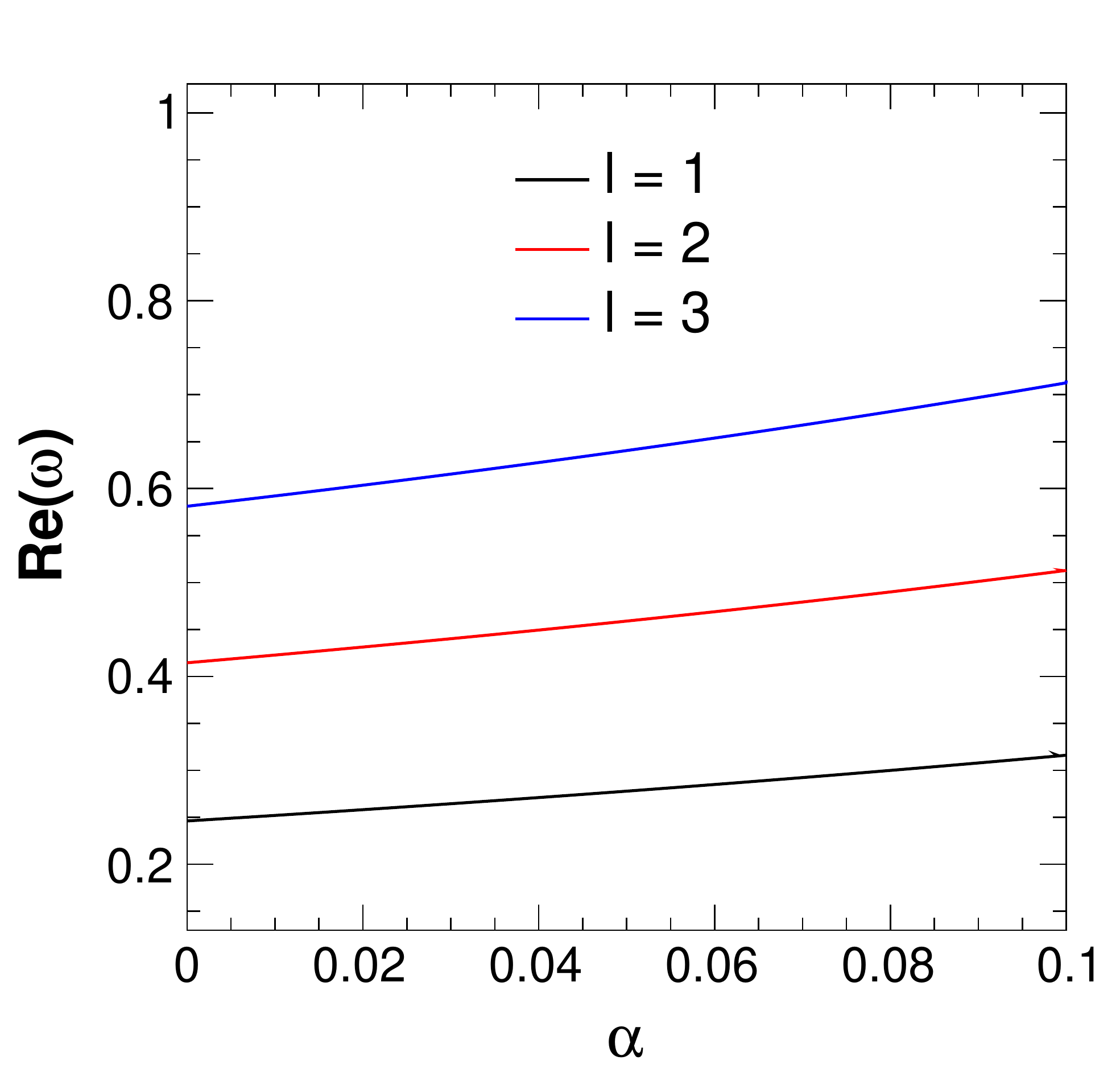}\hspace{0.5cm}
\includegraphics[scale=0.31]{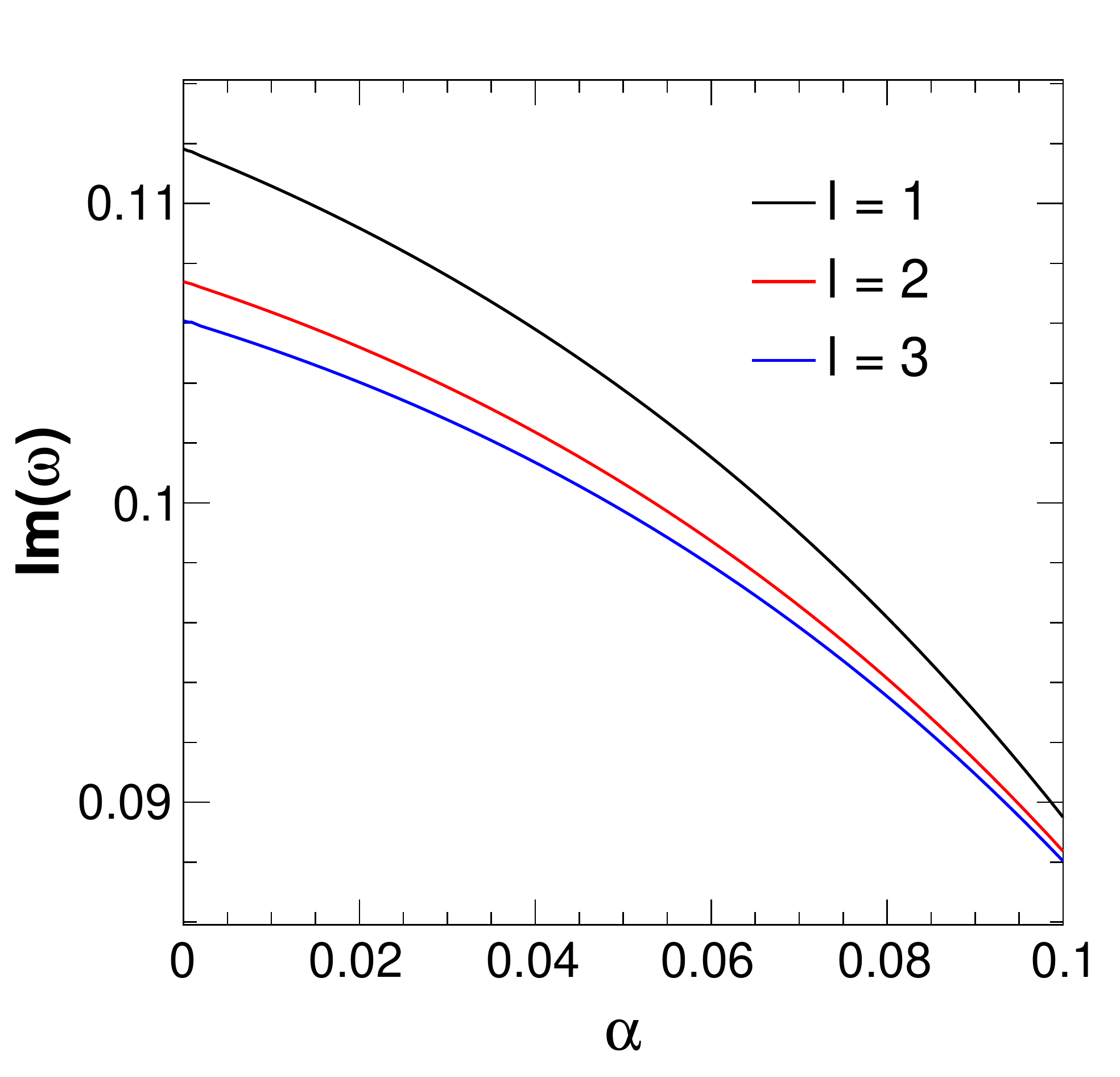}
\caption{Behaviours of QNMs with respect to the GUP parameter $\alpha$ for 
three different values of $l$ with $n=0$ and $\beta=0.05$. The amplitude 
of QNMs increases with an increase in $\alpha$, while the damping decreases 
with increasing $\alpha$ (Mashhoon Method).}
\label{fig2}
\end{figure}

\begin{figure}[h!]
\includegraphics[scale=0.32]{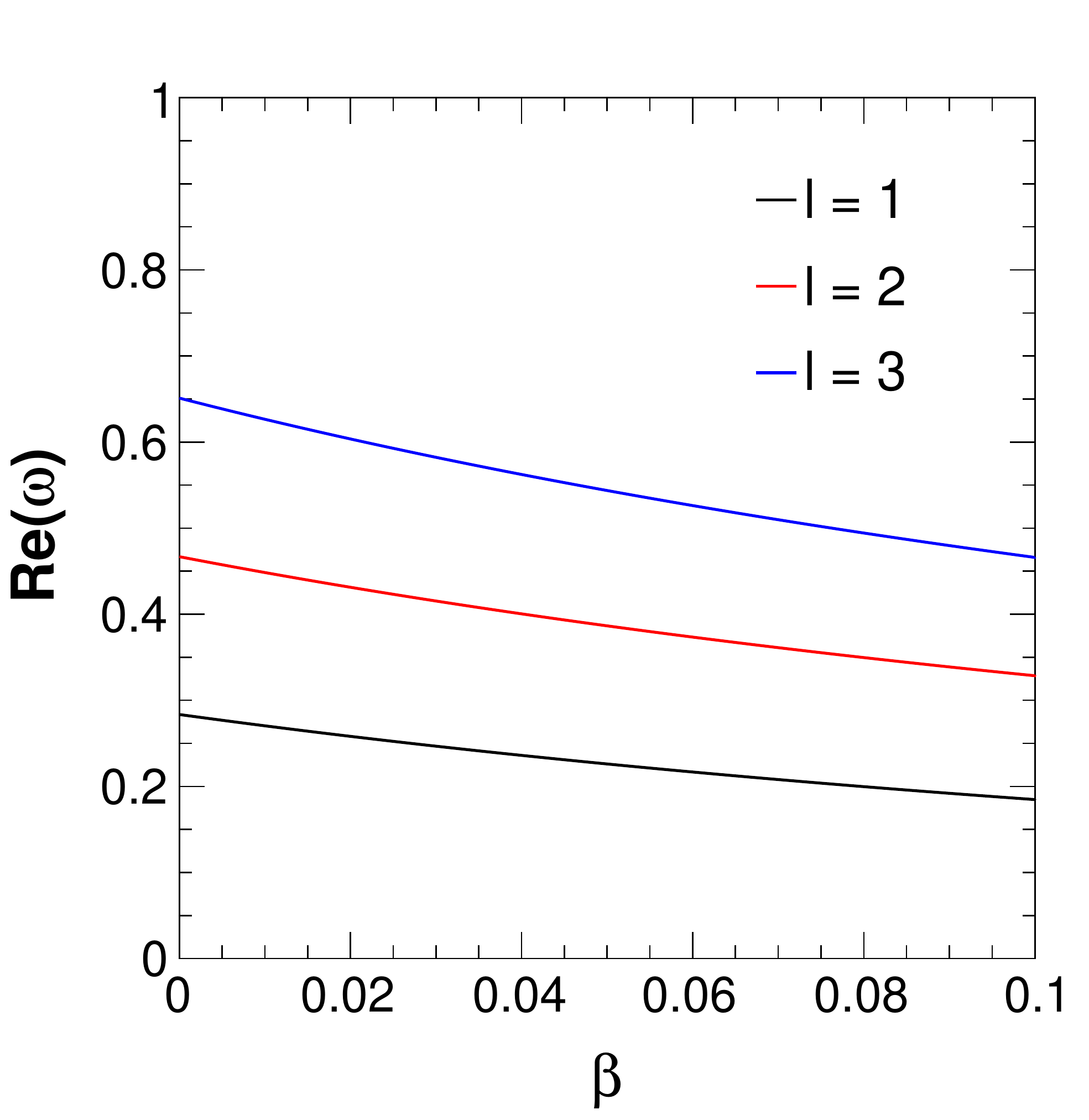}\hspace{0.5cm}
\includegraphics[scale=0.32]{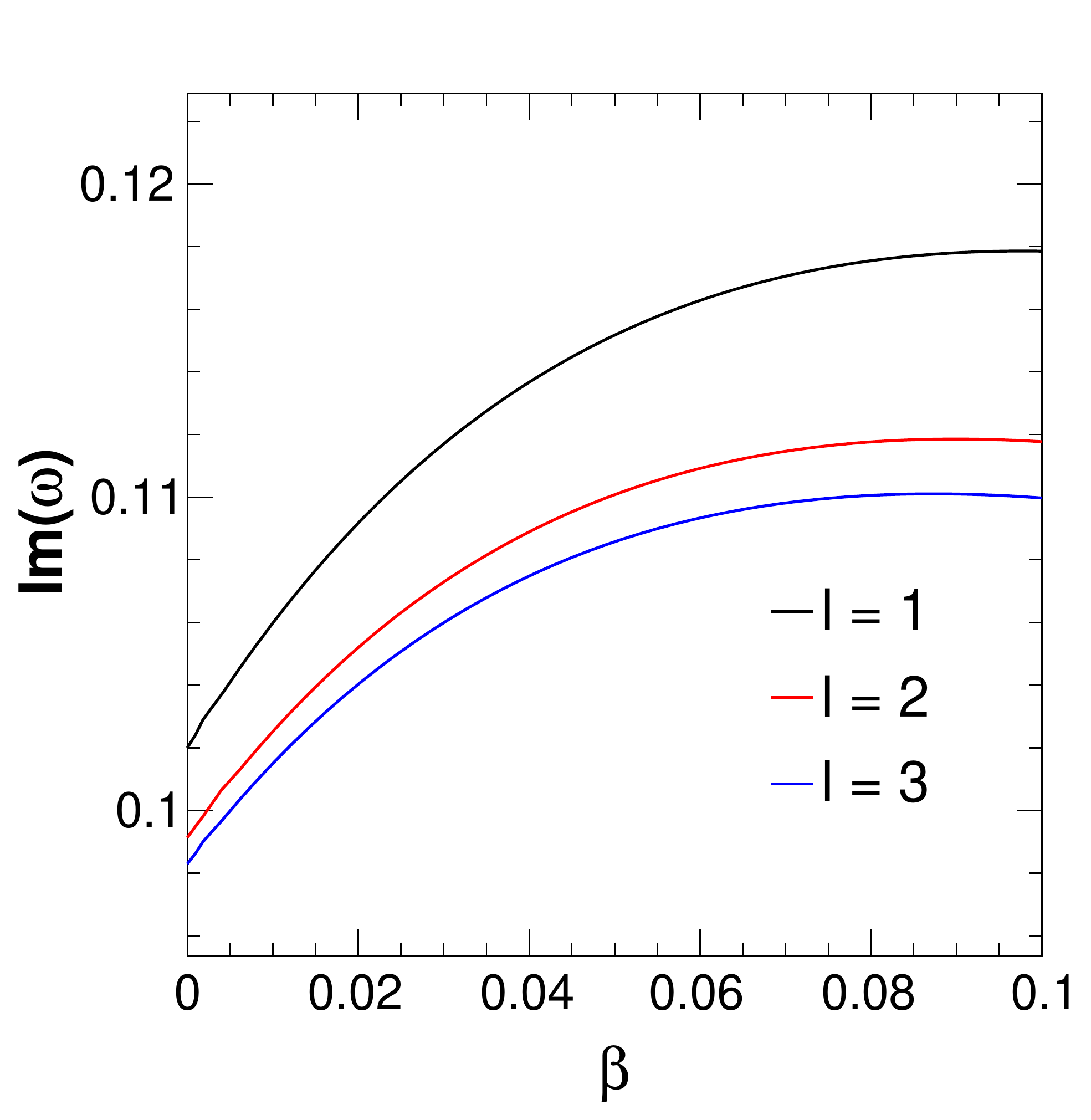}
\caption{Behaviours of QNMs with respect to the GUP parameter $\beta$ for 
three different values of $l$ with $n=0$ and $\alpha=0.05$. The 
amplitude decreases with an increase in $\beta$, while the damping increases 
with increasing $\beta$ (Mashhoon Method).}
\label{fig3}
\end{figure}

\begin{figure}[h!]
\includegraphics[scale=0.28]{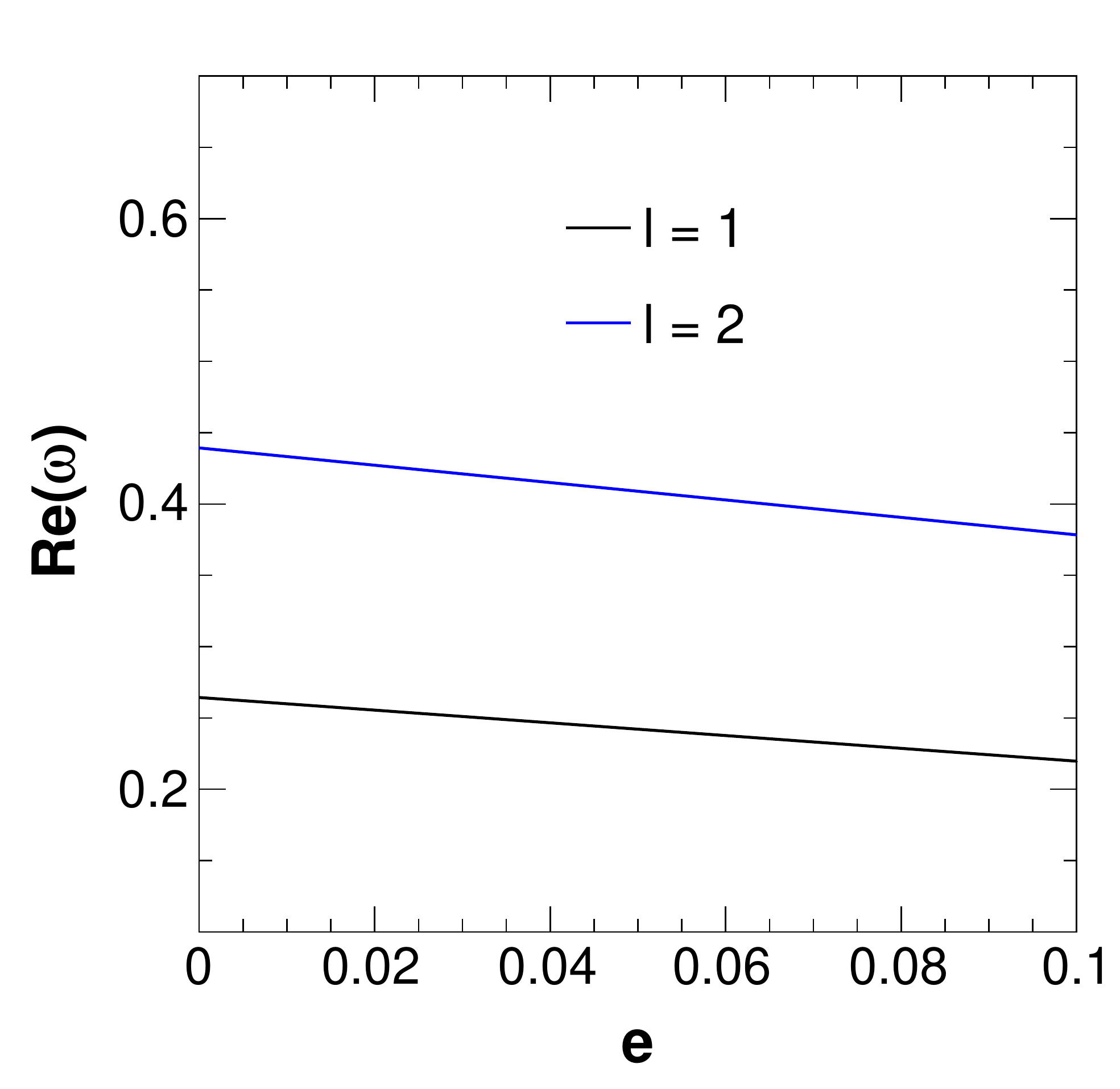}\hspace{0.3cm}
\includegraphics[scale=0.28]{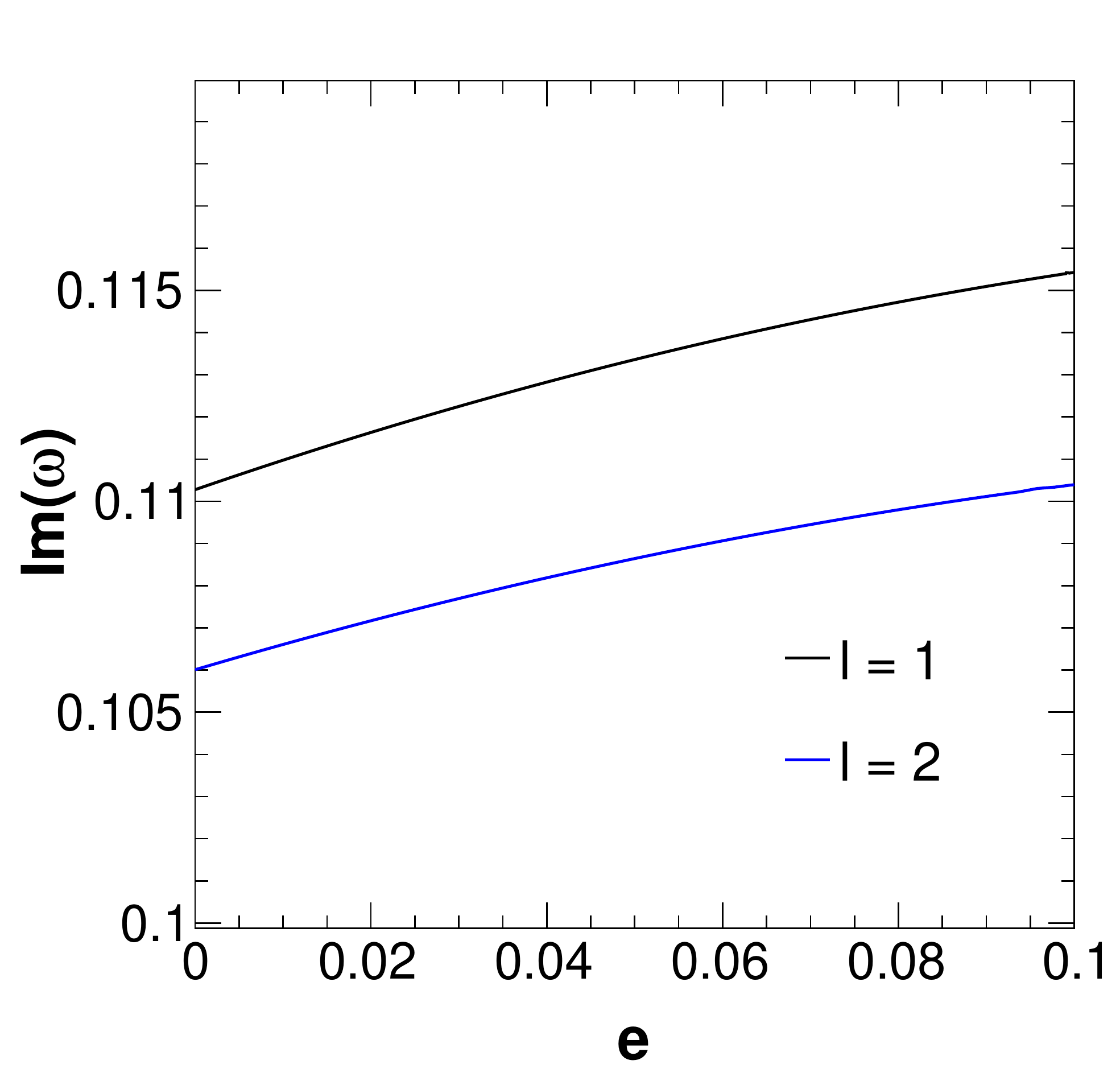}\hspace{0.3cm}
\includegraphics[scale=0.28]{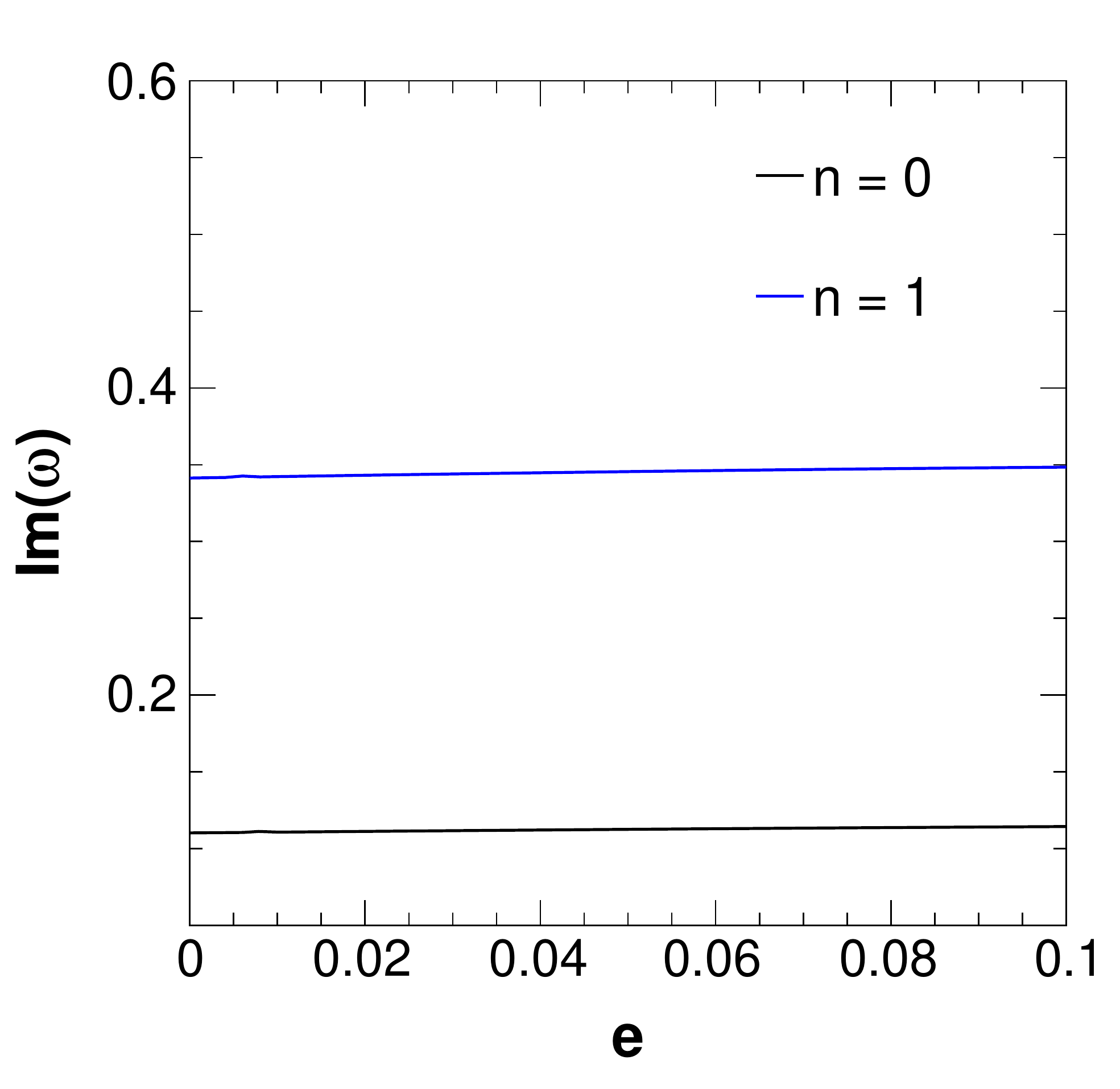}
\caption{Behaviours of QNMs with respect to the quintessence parameter $e$ 
for two different values of $l$ (first two plots with $n=0$) and two different
values $n$ (right plot with $l=1$) with $\alpha=\beta=0.05$. The 
amplitude decreases with an increase in $e$, while the damping increases with 
increasing $e$ (Mashhoon Method).}
\label{fig4}
\end{figure}

\begin{table}[h!]
\caption{QMNs of GUP-corrected black hole surrounded by a 
quintessence field for $n=0$, $n=1$, $n=2$ and $n=3$ modes, for multipole 
number $l=1$ and $l=2$, the deformation parameter $\alpha=0.02$ and 
$\beta=0.02$ and various values of the quintessence parameter $e$ obtained by 
using the Mashhoon method.}
\vspace{2mm}
\centering
\begin{tabular}{c@{\hskip 5pt}c@{\hskip 10pt}c@{\hskip 10pt}c@{\hskip 10pt}c@{\hskip 10pt}c@{\hskip 10pt}c@{\hskip 5pt}c}
\hline \hline
&$l$ & $e$ & QNMs for $n=0$ & QNMs for $n=1$ & QNMs for $n=2$ & QNMs for $n=3$
& \\
\hline \hline
&1 & 0.01 & 0.278884 + 0.106046i & 0.278884 + 0.318139i & 0.278884 + 0.530231i & 0.278884 + 0.742323i &\\
&1 & 0.03 & 0.268485 + 0.107679i & 0.268485 + 0.323037i & 0.268485 + 0.538395i & 0.268485 + 0.753753i &\\
&1 & 0.05 & 0.258121 + 0.109175i & 0.258121 + 0.327526i & 0.258121 + 0.545876i & 0.258121 + 0.764227i &\\
&1 & 0.07 & 0.247793 + 0.110532i & 0.247793 + 0.331597i & 0.247793 + 0.552662i & 0.247793 + 0.773727i &\\
\hline 
&2 & 0.01 & 0.459854 + 0.102496i & 0.459854 + 0.307487i & 0.459854 + 0.512479i & 0.459854 + 0.717470i &\\
&2 & 0.03 & 0.445542 + 0.103907i & 0.445542 + 0.311721i & 0.445542 + 0.519535i & 0.445542 + 0.727349i &\\
&2 & 0.05 & 0.431299 + 0.105201i & 0.431299 + 0.315602i & 0.431299 + 0.526004i & 0.431299 + 0.736406i &\\
&2 & 0.07 & 0.417129 + 0.106375i & 0.417129 + 0.319124i & 0.417129 + 0.531873i & 0.417129 + 0.744623i & \\ 
\hline
\end{tabular}
\label{table03}
\end{table}
Figs.~\ref{fig2} -- \ref{fig4} give the visual representation of all these 
behaviours of QNMs of the black hole as discussed and presented in Tables 
\ref{table01} -- \ref{table03}. Fig.~\ref{fig2} shows the variation of the 
amplitude and damping part of the QNMs with respect to the GUP parameter 
$\alpha$ for a fixed value of the parameter $\beta = 0.05$ and for three 
different $l$ values. It is seen that amplitude of QNMs increases slowly,
whereas the damping decreases rapidly with the increasing values of
$\alpha$. In Fig.~\ref{fig3} the variations of amplitude and damping of QNMs
with respect to $\beta$ for a fixed value of $\alpha = 0.05$ and for three
different $l$ values are shown. This figure shows that the effect of $\beta$
is totally opposite to that of $\alpha$ on the QNMs. However, the trend of the
effect of these two parameters on QNMs is almost similar. The first two plots
of Fig.~\ref{fig4} show the behaviours of amplitude and damping respectively
of QNMs of the black hole with respect to quintessence field parameter $e$ for
a constant value of $\alpha = \beta = 0.05$ and for two values $l$. Similar to
the case of the parameter $\beta$ in this case also the amplitude decreases 
slowly, but the damping increases at a relatively faster step. The third plot 
of this figure shows the fact that for a fixed $l$, the damping almost remains
constant with increasing quintessence parameter $e$ and the higher values of 
$n$ give a much higher damping.

\subsection{QNMs by WKB method}
The QNMs can be reliably calculated using the 6th order WKB method. It is a 
semi-analytical approximation method. The basics of the WKB
method can be found extensively in literature (\cite{konoplya_1,konoplya_2,djgogoi_2} and 
references therein). Here our basic intention is to make a 
comparative analysis of the QNM frequencies obtained by the Mashhoon method
with that will be given by the WKB method. 

\begin{figure}[h!]
\includegraphics[scale=0.28]{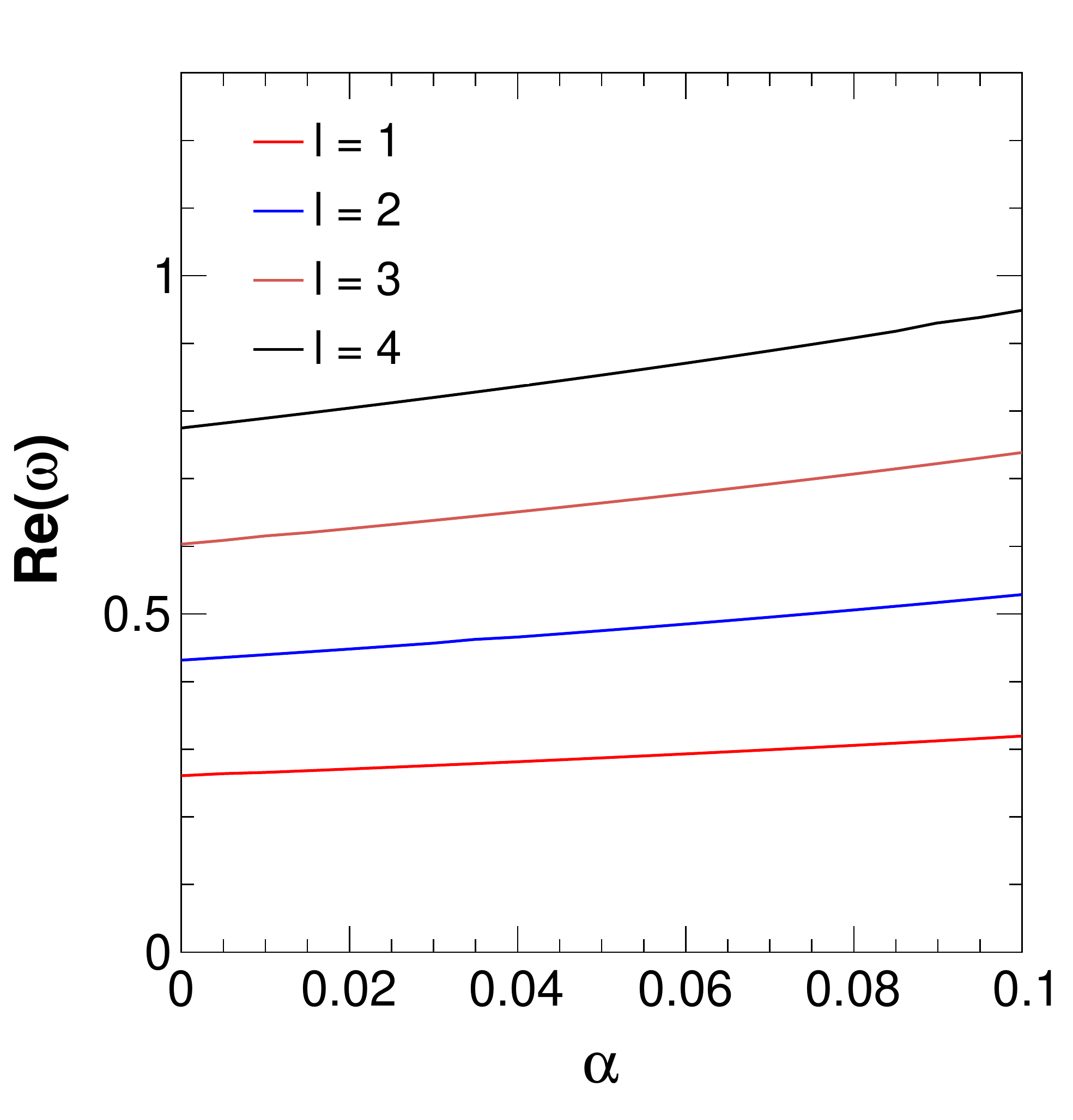}\hspace{0.3cm}
\includegraphics[scale=0.28]{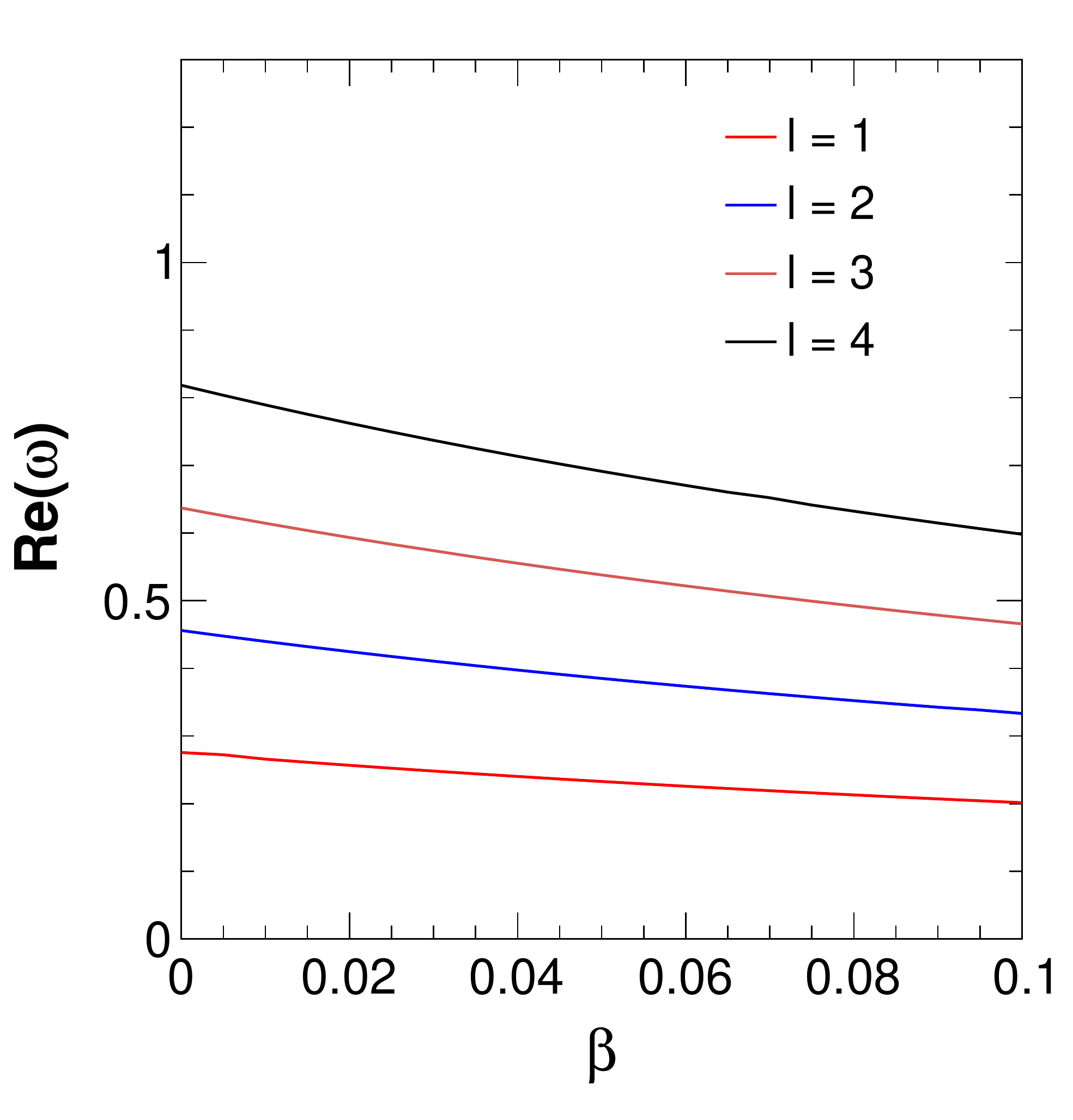}\hspace{0.3cm}
\includegraphics[scale=0.28]{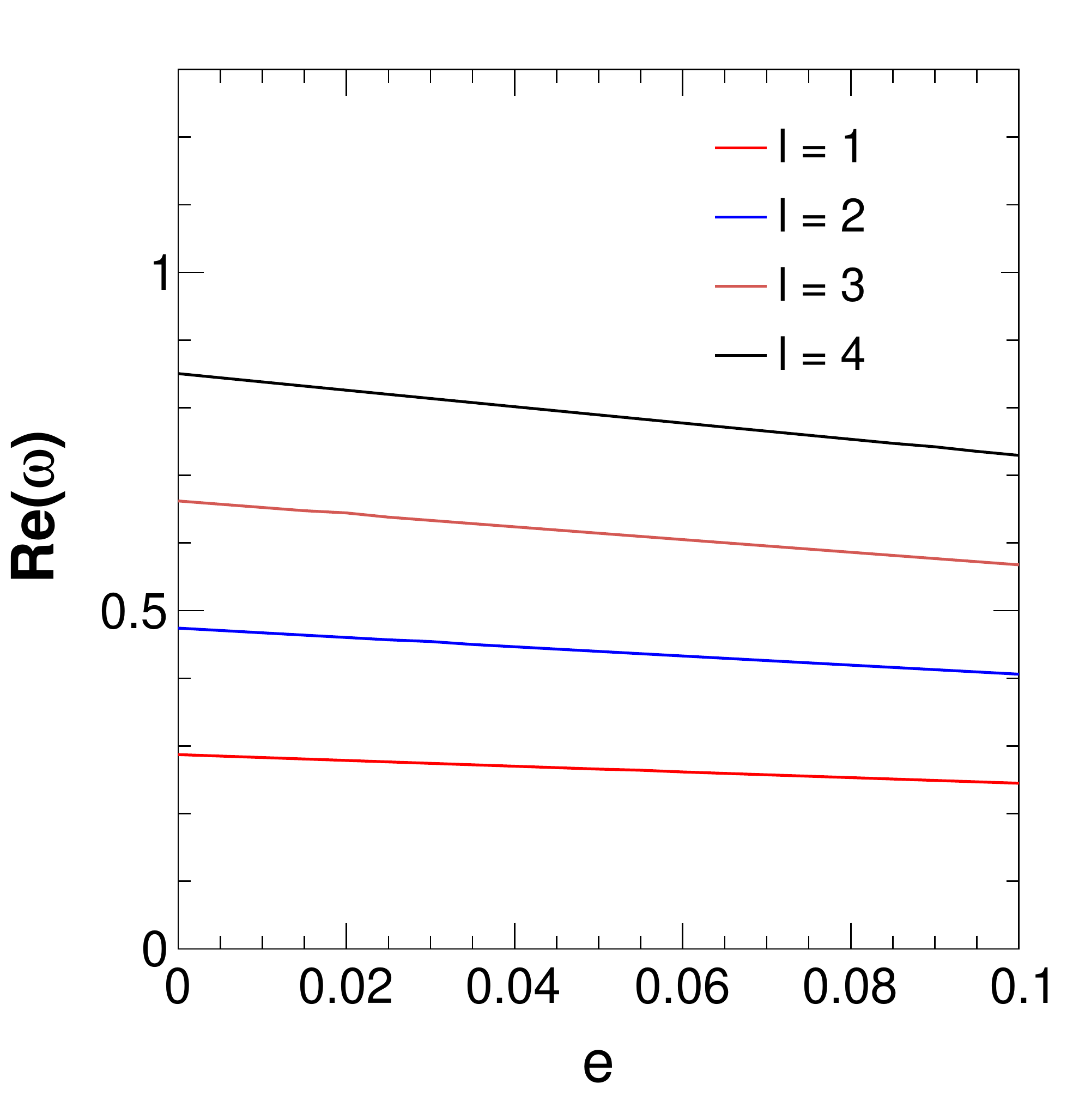}
\caption{Variation of amplitude of QNMs with various parameters
of the model obtained by using the 6th order WKB approximation. The left plot is for
$\beta=0.01$ and $e=0.05$, the middle plot is for $\alpha=0.01$ and $e=0.05$,
and the right plot is for $\alpha=\beta=0.01$.}
\label{fig5}
\end{figure}
\begin{figure}[h!]
\includegraphics[scale=0.28]{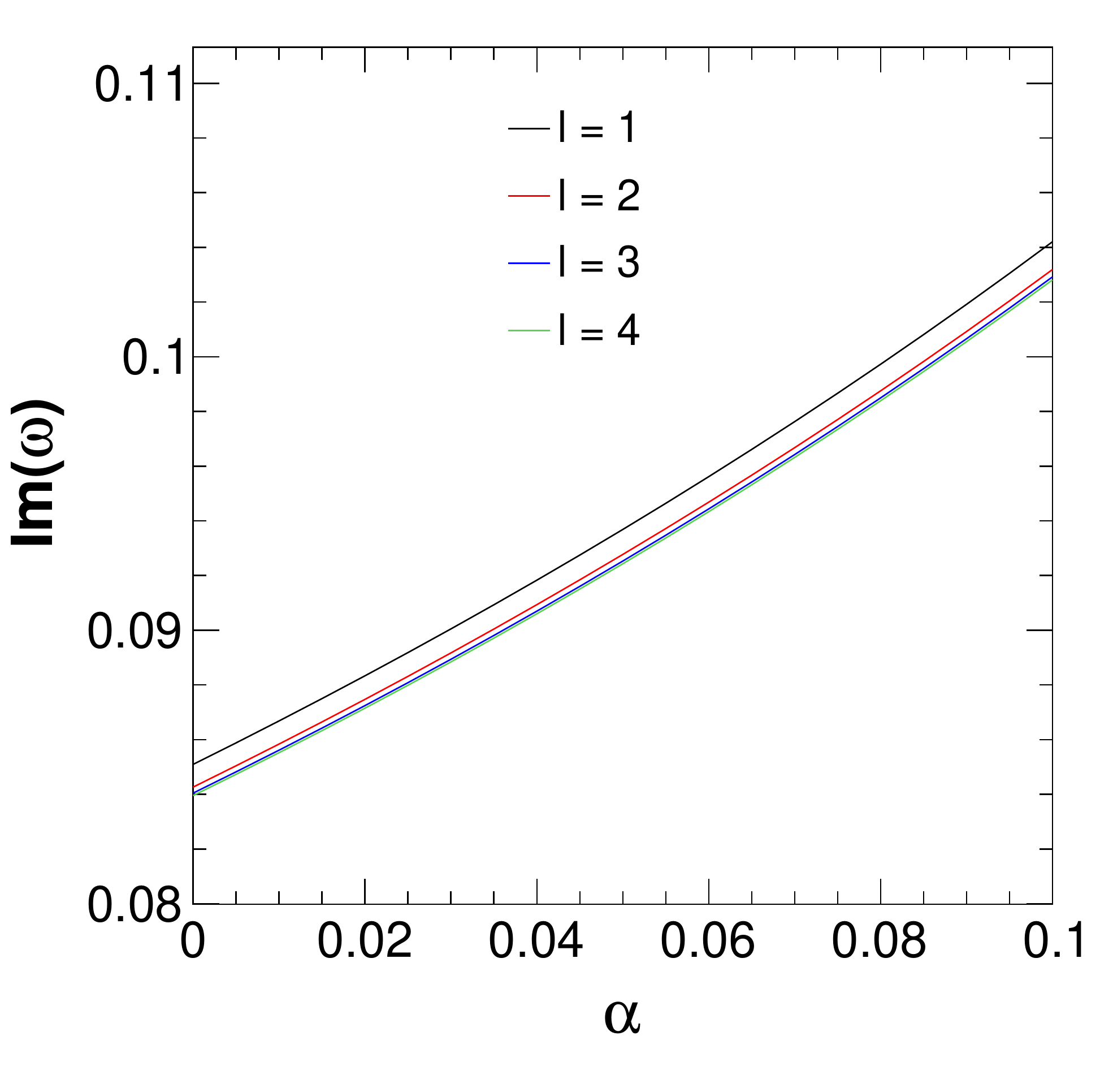}\hspace{0.3cm}
\includegraphics[scale=0.28]{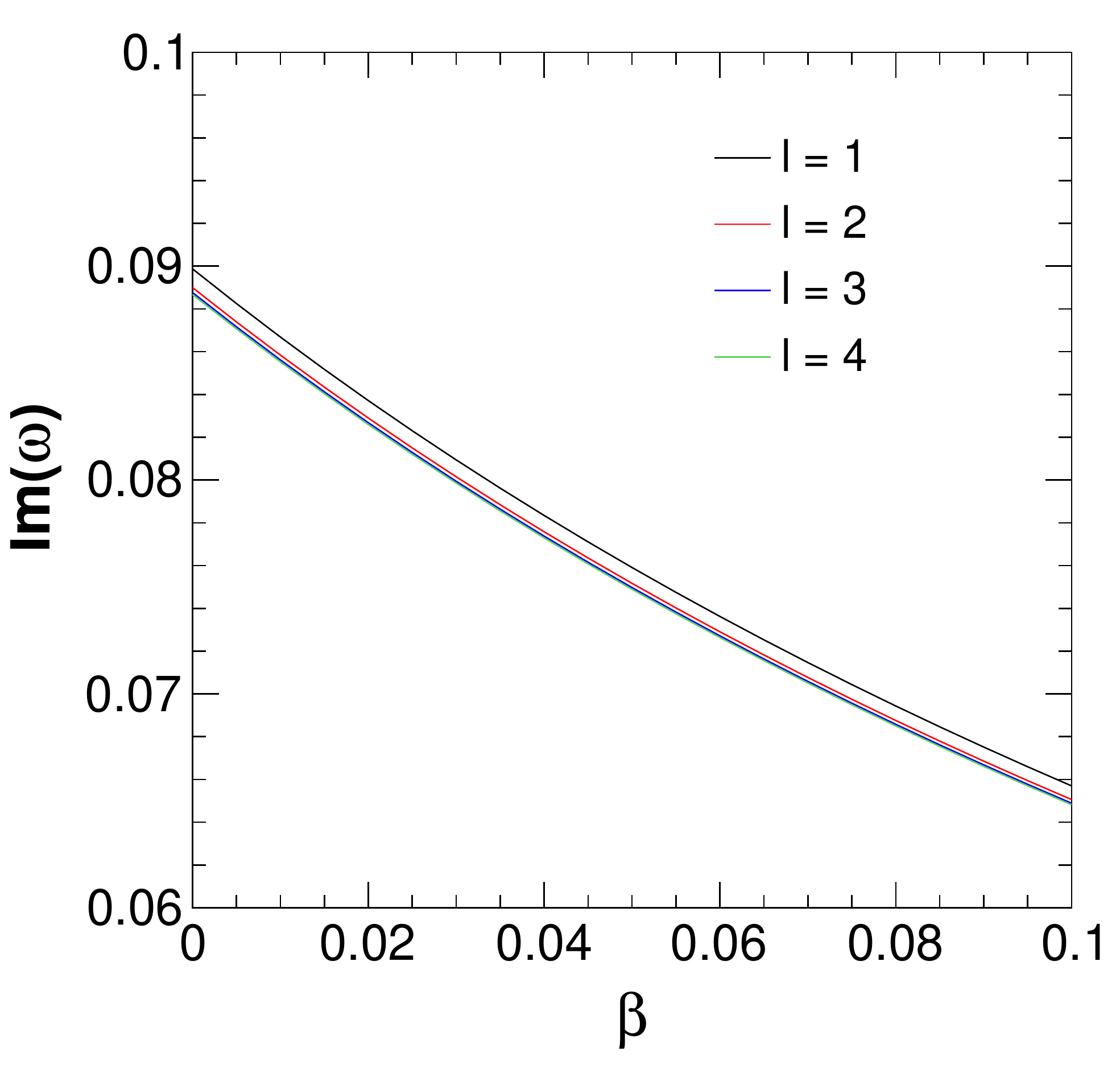}\hspace{0.3cm}
\includegraphics[scale=0.28]{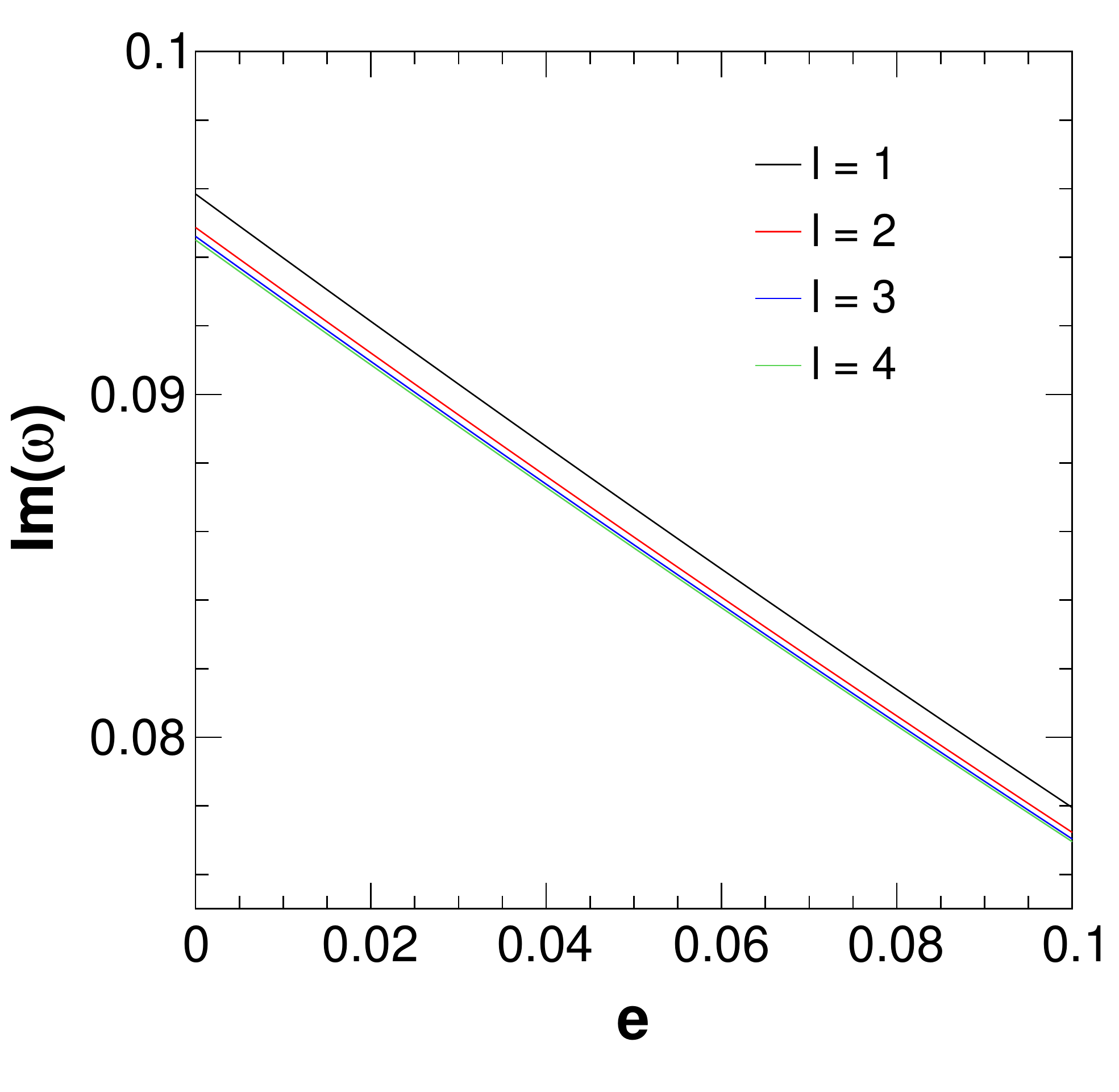}
\caption{Variation of damping of QNMs with various parameters of the model
obtained by using the 6th order WKB approximation. The left plot is for $\beta=0.01$
and $e=0.05$, the middle plot is for $\alpha=0.01$ and $e=0.05$, and the
right plot is for $\alpha=\beta=0.01$.}
\label{fig5ad}
\end{figure}

For consistency, the effective potential of the black hole, represented by equation \eqref{eq12} should satisfy some boundary conditions 
at the horizon and at infinity. Asymptotically flat spacetimes lead to the 
following quasinormal criteria:
\begin{align}
\psi(x) \rightarrow \Bigg \{ \begin{array}{lr}
P e^{+i\omega x} & \text{if }x \rightarrow -\infty \\
Q e^{-i\omega x} & \;\text{if }x \rightarrow +\infty, \end{array}
\label{eq15}
\end{align}
where $P$ and $Q$ denote constants of integration. Using these conditions we 
have calculated the QNM frequencies for our considered black hole. 
The results, i.e.\ the amplitude and damping of QNMs or the real and imaginary
parts of QNM frequencies have been plotted against the GUP and the quintessence 
parameters as shown in Figs.~\ref{fig5} and \ref{fig5ad}. Fig.~\ref{fig5} 
shows the trends of variations of the real parts of the QNMs with respect to 
variations in $\alpha$, $\beta$ and $e$. It is clear that these trends are the 
same with the amplitude obtained by the Mashhoon method, but in this case 
the value of the amplitude is slightly greater than that obtained by the 
Mashhoon method. Whereas the damping of the QNMs shows the opposite behaviour 
with the Mashhoon method. However, it is to be noted that in WKB method, the 
variation of damping term is insignificant with respect to variation in 
all the model parameters.   

Tables \ref{table04} and \ref{table05} show a clear comparison between the 
Mashhoon method and the 6th order WKB method. Table \ref{table04} is for the $n=0$ case
and Table \ref{table05} is for the $n=1$. For $n=0$ we found that 
the real part of the QNM calculated by the two methods agree to a good 
extent while the imaginary part of the modes vary to some extent. There is a 
far better agreement between two methods for higher $l$ values. From Table 
\ref{table04} one can observe the following additional points. For the given 
values of $l$, $\beta$ and $e$ with increase in $\alpha$ values, there is an 
increasing mismatch between the two methods. On the other hand, for the given 
values of $l$, $\alpha$ and $e$ with increasing values of $\beta$ there is a 
better match between the methods. Moreover, for the given values of $l$, 
$\alpha$ and $\beta$, increasing $e$ values lead to decrease in mismatch 
between the methods. However, from Table \ref{table05} it is seen that some
of the above observations are not applicable to the case for $n=1$ and the 
difference between the Mashhoon method and the 6th order WKB method in this case is 
almost ten times higher than that for the $n=0$ case. This may be due to the 
fact that the usual WKB method can only be reliably used to calculate QNMs for the 
situation with $n < l$ \cite{konoplya_1,djgogoi_2}. 

At this point we would like to comment that the deviation between 
results obtained from the two methods is obvious as both are approximation 
methods. Till now there is no experimental data on QNMs to validate which 
method is more accurate than the other. Notwithstanding, in literature we 
can see that WKB method is the most common choice for computing QNMs as this 
method has a systematic procedure for computing errors and improving accuracy 
by considering higher order corrections.

\begin{table}[h!]
\caption{Comparison of QNMs of GUP-corrected black hole 
surrounded by a quintessence field obtained by using the Mashhoon method 
and the 6th order WKB method for $n=0$, $l=1,2,3,4$, and for various values 
of the parameters $\alpha$, $\beta$ and $e$.
Here $\Delta |\omega_M-\omega_{WKB}|$ represents the absolute difference
between the QNM frequencies calculated by using the Mashhoon method and the 6th order WKB 
method.}
\vspace{2mm}
\centering
\begin{tabular}{c@{\hskip 5pt}c@{\hskip 10pt}c@{\hskip 10pt}c@{\hskip 10pt}c@{\hskip 10pt}c@{\hskip 10pt}c@{\hskip 10pt}c@{\hskip 5pt}c}
\hline \hline
&Multipole & $\alpha$ & $\beta$ & $e$ & Mashhoon Method & 6th order WKB method & $\Delta |\omega_M-\omega_{WKB}|$ & \\
\hline
&\multirow{4}{4em}{$l=1$} & $0.01$ & $0.01$ & $0.03$ & 0.274715 + 0.105993i & 0.274250 + 0.090298i & $5.7204 \times 10^{-3}$ &\\ 
&& $0.01$ & $0.01$ & $0.05$ & 0.263814 + 0.107754i & 0.265742 + 0.086685i & $5.4485 \times 10^{-3}$ &\\
&& $0.01$ & $0.03$ & $0.05$ & 0.240990 + 0.112803i & 0.248128 + 0.080939i & $5.0885 \times 10^{-3}$ &\\
&& $0.05$ & $0.01$ & $0.05$ & 0.291550 + 0.099263i & 0.287202 + 0.093685i & $5.8888 \times 10^{-3}$ &\\
\hline
&\multirow{4}{4em}{$l=2$} & $0.01$ & $0.01$ & $0.03$ & 0.454274 + 0.102489i & 0.453426 + 0.089402i & $3.5361 \times 10^{-3}$ & \\
&& $0.01$ & $0.01$ & $0.05$ & 0.439261 + 0.104011i & 0.439746 + 0.085840i & $3.3614 \times 10^{-3}$ &\\
&& $0.01$ & $0.03$ & $0.05$ & 0.407365 + 0.108192i & 0.410599 + 0.080150i & $3.1389 \times 10^{-3}$ &\\
&& $0.05$ & $0.01$ & $0.05$ & 0.478163 + 0.096792i & 0.475258 + 0.092772i & $3.6331 \times 10^{-3}$ & \\
\hline
&\multirow{4}{4em}{$l=3$} & $0.01$ & $0.01$ & $0.03$ & 0.634124 + 0.101454i & 0.633405 + 0.089164i & $2.5386 \times 10^{-3}$ &\\ 
&& $0.01$ & $0.01$ & $0.05$ & 0.614233 + 0.102908i & 0.614444 + 0.085616i & $2.4137 \times 10^{-3}$ &\\
&& $0.01$ & $0.03$ & $0.05$ & 0.571617 + 0.106832i & 0.573718 + 0.079941i & $2.2538 \times 10^{-3}$ &\\
&& $0.05$ & $0.01$ & $0.05$ & 0.666197 + 0.096067i & 0.664065 + 0.092530i & $2.6074 \times 10^{-3}$ & \\
\hline
&\multirow{4}{4em}{$l=4$} & $0.01$ & $0.01$ & $0.03$ & 0.814241 + 0.101021i & 0.813646 + 0.089066i & $1.9775 \times 10^{-3}$ & \\ 
&& $0.01$ & $0.01$ & $0.05$ & 0.789248 + 0.102448i & 0.789370 + 0.085524i & $1.8798 \times 10^{-3}$ &\\
&& $0.01$ & $0.03$ & $0.05$ & 0.735481 + 0.106263i & 0.737049 + 0.079855i & $1.7553 \times 10^{-3}$ &\\
&& $0.05$ & $0.01$ & $0.05$ & 0.854793 + 0.095764i & 0.853116 + 0.092430i & $2.0321 \times 10^{-3}$ &\\
\hline
\end{tabular}
\label{table04}
\end{table}

\begin{table}[h!]
\caption{Comparison of QNMs of GUP-corrected black hole
surrounded by a quintessence field obtained by using the Mashhoon method
and the 6th order WKB method for $n=1$, $l=1,2,3,4$, and for various values 
of the parameters $\alpha$, $\beta$ and $e$.
Here $\Delta |\omega_M-\omega_{WKB}|$ represents the absolute difference 
between the QNM frequencies calculated by using the Mashhoon method and the 6th order 
WKB method.}
\vspace{2mm}
\centering
\begin{tabular}{c@{\hskip 5pt}c@{\hskip 10pt}c@{\hskip 10pt}c@{\hskip 10pt}c@{\hskip 10pt}c@{\hskip 10pt}c@{\hskip 10pt}c@{\hskip 5pt}c}
\hline \hline
&Multipole & $\alpha$ & $\beta$ & $e$ & Mashhoon Method & 6th order WKB method & $\Delta |\omega_M-\omega_{WKB}|$ & \\
\hline
& \multirow{4}{4em}{$l=1$} & $0.01$ & $0.01$ & $0.03$ & 0.274715 + 0.317980i & 0.248160 + 0.282859i & $4.3926 \times 10^{-2}$ & \\
&& $0.01$ & $0.01$ & $0.05$ & 0.263814 + 0.323261i & 0.240814 + 0.271373i & $5.4432 \times 10^{-2}$ & \\
&& $0.01$ & $0.03$ & $0.05$ & 0.240990 + 0.338408i & 0.224852 + 0.253387i & $7.6679 \times 10^{-2}$ & \\
&& $0.05$ & $0.01$ & $0.05$ & 0.291550 + 0.297790i & 0.260262 + 0.293288i & $2.4635 \times 10^{-2}$ &  \\
\hline
& \multirow{4}{4em}{$l=2$} & $0.01$ & $0.01$ & $0.03$ & 0.454274 + 0.307466i & 0.435359 + 0.272995i & $3.4672 \times 10^{-2}$ & \\
&& $0.01$ & $0.01$ & $0.05$ & 0.439261 + 0.312032i & 0.422543 + 0.262032i & $4.1613 \times 10^{-2}$ & \\
&& $0.01$ & $0.03$ & $0.05$ & 0.407365 + 0.324575i & 0.394536 + 0.244664i & $5.6619 \times 10^{-2}$  & \\
&& $0.05$ & $0.01$ & $0.05$ & 0.478163 + 0.290375i & 0.456666 + 0.283193i & $2.2079 \times 10^{-2}$  & \\
\hline
&\multirow{4}{4em}{$l=3$} & $0.01$ & $0.01$ & $0.03$ & 0.634124 + 0.304361i & 0.620011 + 0.269992i & $2.7137 \times 10^{-2}$ & \\
&& $0.01$ & $0.01$ & $0.05$ & 0.614233 + 0.308725i & 0.601704 + 0.259200i & $3.2296 \times 10^{-2}$ & \\
&& $0.01$ & $0.03$ & $0.05$ & 0.571617 + 0.320496i & 0.561822 + 0.242020i & $4.3601 \times 10^{-2}$ & \\
&& $0.05$ & $0.01$ & $0.05$ & 0.666197 + 0.288200i & 0.650296 + 0.280132i & $1.7796 \times 10^{-2}$ & \\
\hline
&\multirow{4}{4em}{$l=4$} & $0.01$ & $0.01$ & $0.03$ & 0.814241 + 0.303063i & 0.803072 + 0.268725i & $2.1973 \times 10^{-2}$ & \\
&& $0.01$ & $0.01$ & $0.05$ & 0.789248 + 0.307343i & 0.779316 + 0.258007i & $2.6063 \times 10^{-2}$ & \\
&& $0.01$ & $0.03$ & $0.05$ & 0.735481 + 0.318789i & 0.727662 + 0.240906i & $3.5094 \times 10^{-2}$ & \\
&& $0.05$ & $0.01$ & $0.05$ & 0.854793 + 0.287293i & 0.842251 + 0.278843i & $1.4571 \times 10^{-2}$ & \\
\hline
\end{tabular}
\label{table05}
\end{table}
                                
\section{Thermodynamic Properties of the Black hole}
The notion of a minimal length scale does not exist in the normal 
Heisenberg algebra, but in the Planck energy scales, taking into 
consideration the effects of gravity, it becomes necessary. The 
introduction of the GUP is thus naturally motivated and leads to 
interesting results. The temperature of the Schwarzschild black hole 
is usually expressed in the form \cite{liang}:
\begin{equation}
T=\frac{\kappa}{8\pi}\frac{dA}{dS},
\label{eq16}
\end{equation}
where $\kappa$ is the surface gravity of the black hole, $A$ is the 
surface area and $S$ is the entropy of the black hole. Calculations yield the 
expression for the surface gravity at the horizon in our case as
\begin{equation}
\kappa=-\lim_{r\to r_H}\sqrt{-\frac{g^{11}}{g^{00}}}\,\frac{(g^{00})^{'}}{g^{00}}=\frac{1}{r_H}\Big(1+\frac{3e \omega}{r_H^{3\omega+1}}\Big),
\label{eq17}
\end{equation}
where the GUP corrected horizon radius $r_{hGUP}$ of the black hole is denoted
as $r_H$. Liang \cite{liang} showed that the area of a black hole 
increases proportionately when it absorbs a particle of particular mass and 
size. A minimal change in area means a minimal change in entropy, 
which can have the smallest possible value of $\ln2$ according to 
the information theory. So, we can express the ratio $\frac{dA}{dS}$ as
\begin{equation}
\frac{dA}{dS}=\frac{(\Delta A)_{min}}{(\Delta S)_{min}}=\frac{\epsilon \Delta x \Delta p}{\ln2}= \frac{\epsilon (4r_H +\alpha)r_H}{\beta \ln2}\left[1-\sqrt{1-\frac{4\beta}{(4r_H+\alpha)^2}}\right].
\label{eq18}
\end{equation}
Here, the uncertainties in momentum and position are connected with the mass 
and size of the particle falling into the black hole respectively and 
$\epsilon$ is the calibration factor \cite{bcl}. Using equations \eqref{eq17} 
and \eqref{eq18} into \eqref{eq16}, we have the following expression for the 
GUP-corrected temperature,
\begin{equation}
T_{GUP}=\frac{1}{8\pi}\left(1+\frac{3e \omega}{r_H^{3\omega+1}}\right)\frac{\epsilon (4r_H +\alpha)}{\beta \ln2}\left[1-\sqrt{1-\frac{4\beta}{(4r_H+\alpha)^2}}\right].
\label{eq19}
\end{equation}
The above expression of temperature modifies into a simpler form 
in absence of the quintessence and the deformation parameters, which 
must be equal to the Hawking temperature $\frac{1}{4\pi r_H}$. Thus, 
the factor $\epsilon$ is determined to be $4 \ln2$ and we have the 
final form of the GUP-modified black hole temperature as
\begin{equation}
T_{GUP}=\frac{(4r_H +\alpha)}{2\pi\beta }\Big(1+\frac{3e \omega}{r_H^{3\omega+1}}\Big)\left[1-\sqrt{1-\frac{4\beta}{(4r_H+\alpha)^2}}\right].
\label{eq20}
\end{equation} 
For $\omega=-\frac{1}{3}$, the expression for the GUP-corrected temperature 
can be simplified as
\begin{equation}
T_{GUP}= \frac{(4r_H +\alpha)}{2\pi\beta}(1-e)\left[1-\sqrt{1-\frac{4\beta}{(4r_H+\alpha)^2}}\right].
\label{eq21}
\end{equation}

\begin{figure}
\includegraphics[scale=0.28]{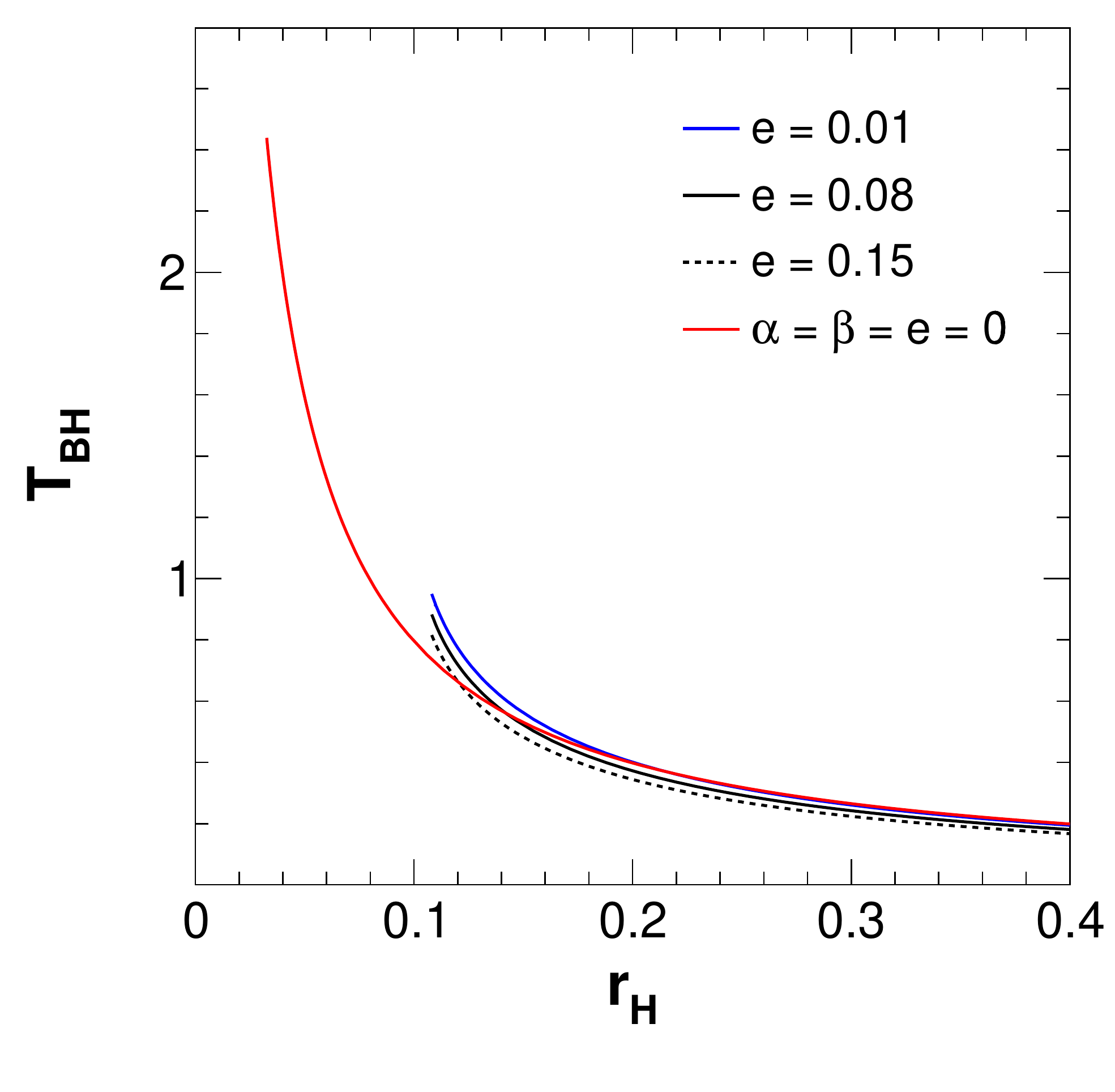}\hspace{0.3cm}
\includegraphics[scale=0.28]{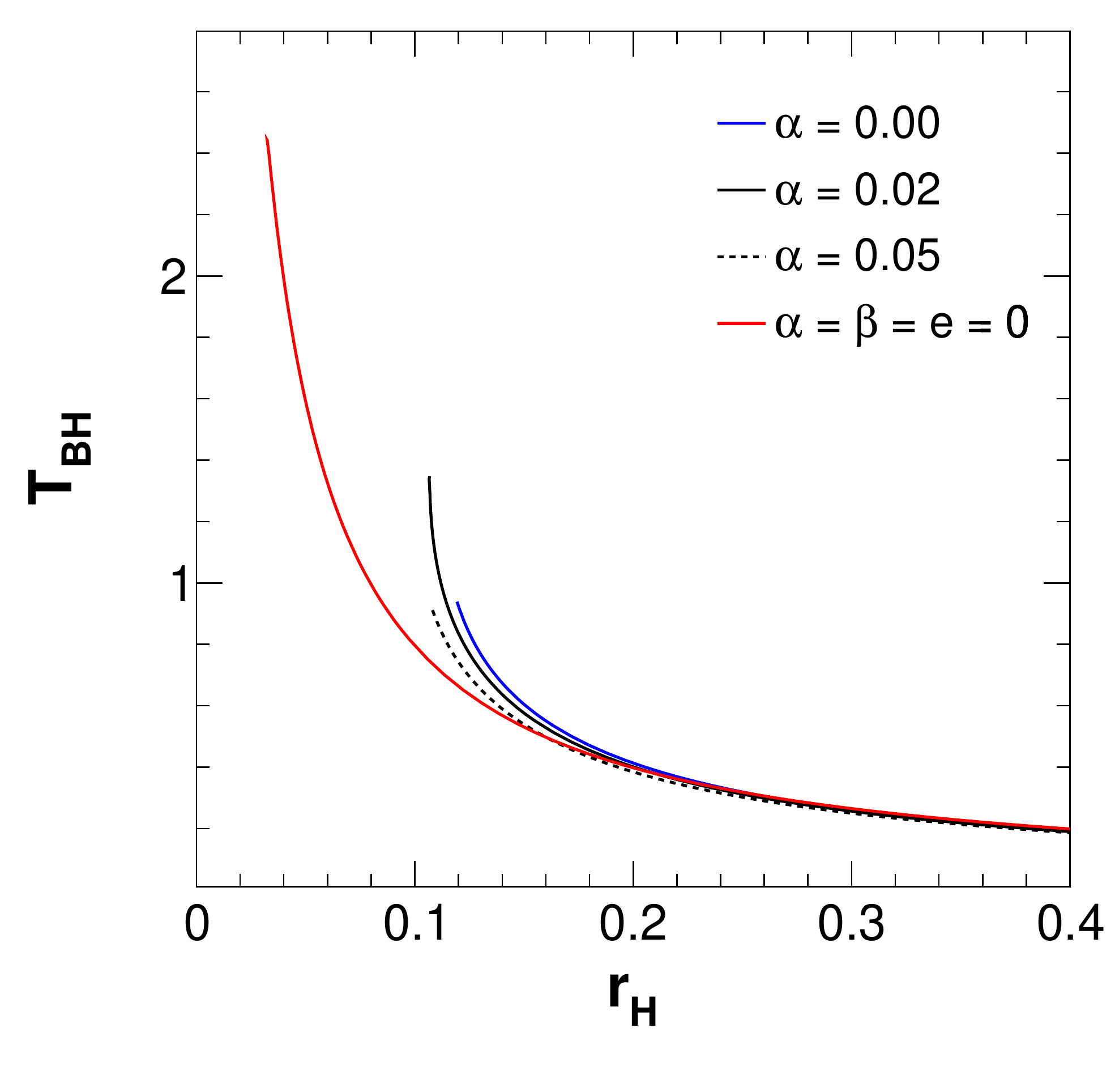}\hspace{0.3cm}
\includegraphics[scale=0.28]{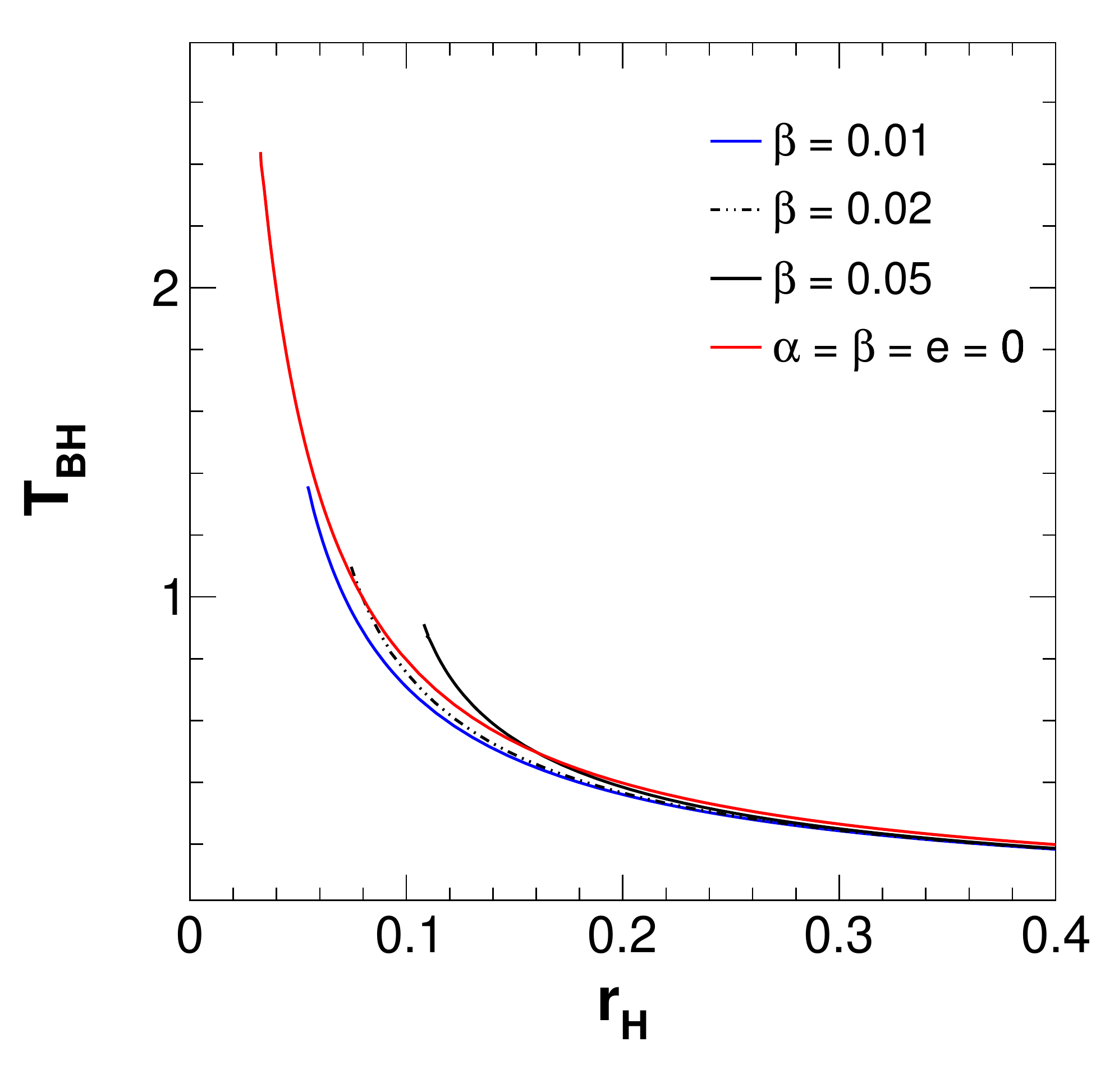}
\caption{Variation of the GUP-corrected black hole temperature with 
quintessence parameter $e$ for $\alpha=\beta=0.05$ (left plot), with $\alpha$ 
considering $\beta= e = 0.05$ (middle plot) and with $\beta$ keeping 
$\alpha = e = 0.05$ (right plot). In all three plots the solid red line 
represents the Schwarzschild black hole temperature.}
\label{fig6}
\end{figure}

It is interesting to note that the introduction of GUP corrections lead to 
the dependency of the temperature on the deformation parameters $\alpha$ 
and $\beta$, apart from the quintessence parameter $e$. In absence of the 
deformation parameters, the HUP-corrected temperature of the Schwarzschild 
black hole surrounded by quintessence is given by
\begin{equation}
T_{HUP}=\frac{1}{4\pi r_H}\Big(1+\frac{3e \omega}{r_H^{3\omega+1}}\Big).
\label{eq22}
\end{equation}
Considering real-valued temperature, it can be seen from the above 
equation \eqref{eq20} that for $\omega=-\frac{1}{3}$, there exists some bounds
on the horizon of the black hole depending on the values of $\alpha$ and 
$\beta$ \cite{bcl}, since the term inside the square root can not be negative. 
This is illustrated by the plots of temperature vs horizon graphs shown in
Fig.~\ref{fig6}, which clearly shows the dependence as mentioned.
From this figure it is clear that introduction of quantum corrections due 
to the generalised uncertainty principle and the surrounding quintessence 
field have some influence on the black hole temperature. In some cases the
influence looks significant for black holes with small event horizon radii.
Whereas for black holes with large event horizon radius influence looks 
insignificant. In all the 
cases, temperature seems to decrease with the increase in horizon radius. Thus, 
black holes with very large event horizon radii might have sub-zero 
temperatures associated with them. The temperature profiles obtained here are 
in good agreement with results presented in \cite{bcl,adler}. 

\begin{figure}
\includegraphics[scale=0.28]{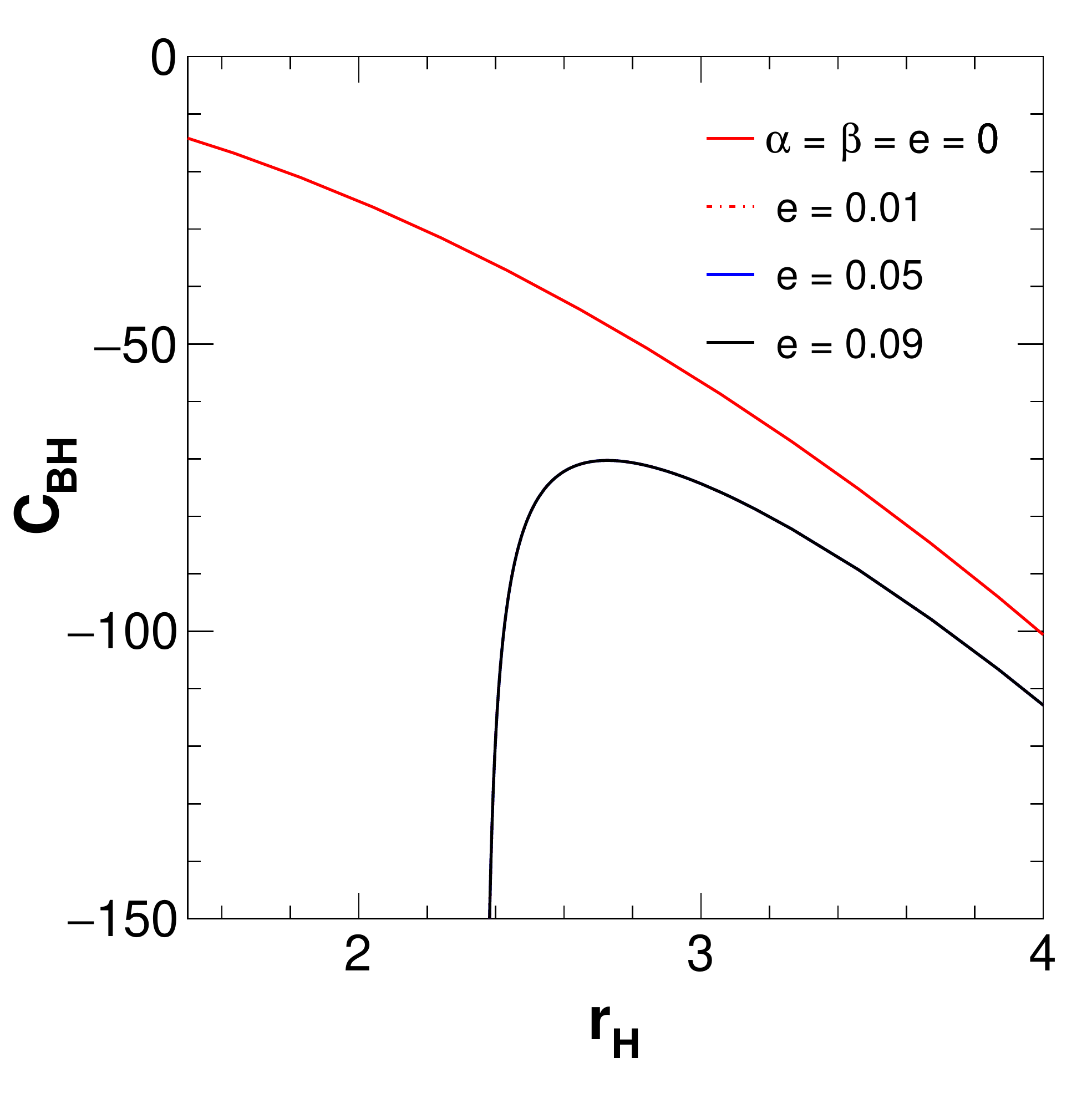}\hspace{0.3cm}
\includegraphics[scale=0.28]{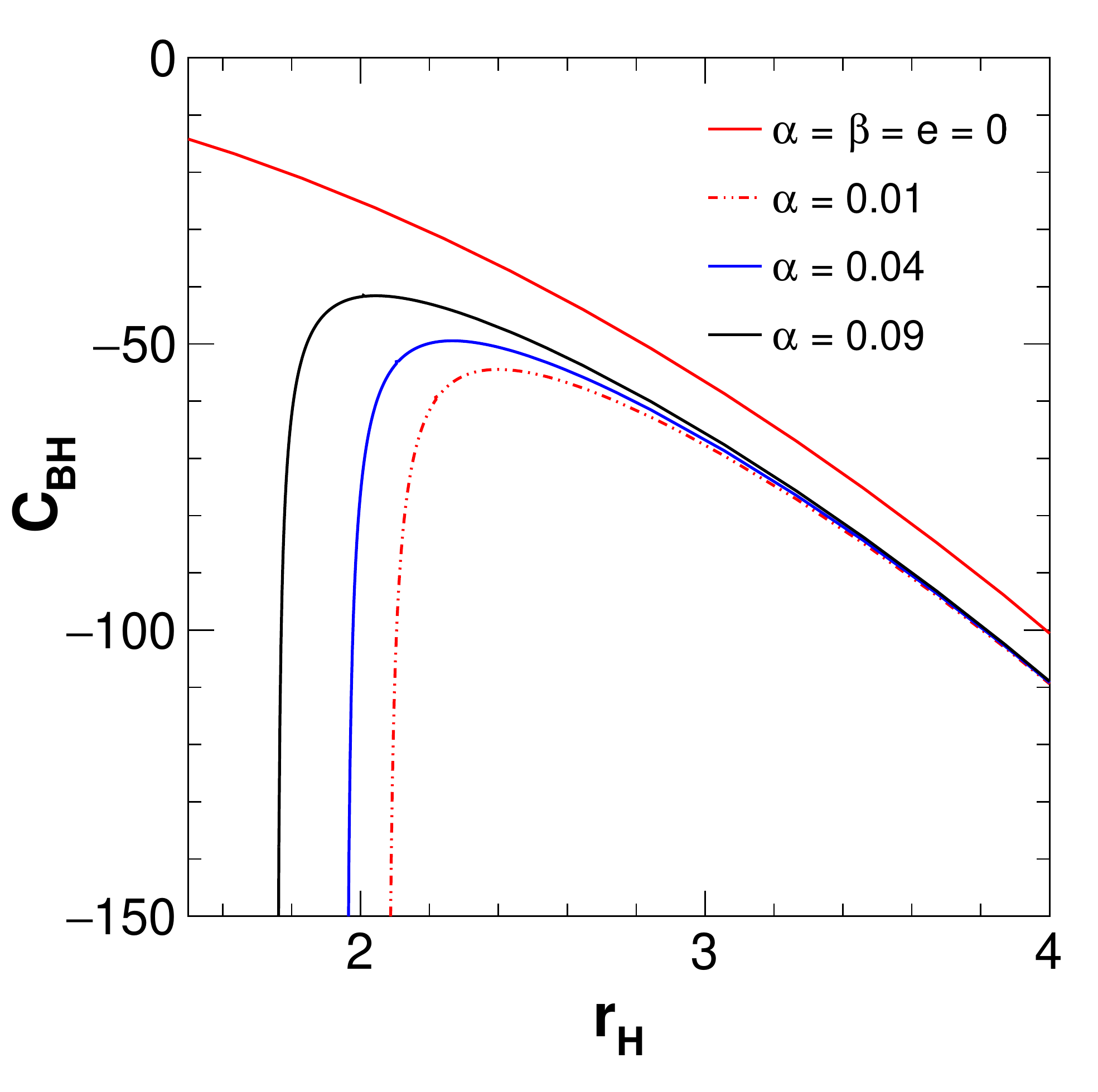}\hspace{0.3cm}
\includegraphics[scale=0.28]{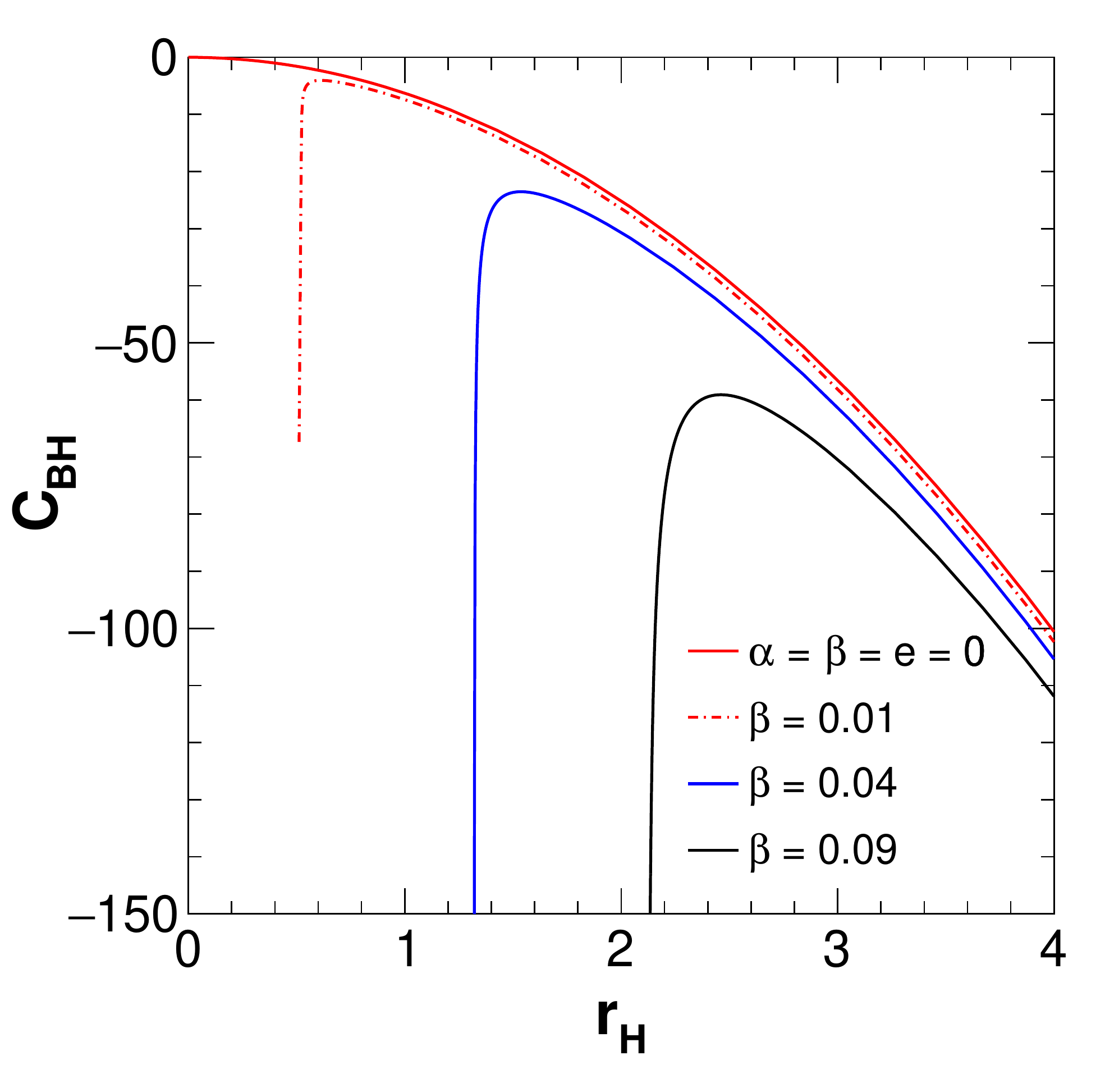}
\caption{Variation of the GUP-corrected heat capacity of the black hole
$C_{BH}$ in terms of horizon radius $r_H$ for varying $e$ values with
$\alpha=0.01$ and $\beta = 0.09$ (left plot), for varying $\alpha$ values with
$\beta = e = 0.07$ (middle plot) and for varying values of $\beta$ with
$\alpha=0.07$ and $e=0.05$ (right plot). In all three plots the solid red line
represents the heat capacity of Schwarzschild black hole.}
\label{fig7}
\end{figure}

Another interesting aspect of black hole physics is the study of the remnant 
formation, which is considered to be a stable state of the black hole that 
does not emit any heat and whose mass is reduced due to the evaporation. In this
context the dependency of heat capacity of the black hole on the GUP 
parameters becomes an important feature that we want to analyse. For this 
purpose, we make use of the thermodynamic relation connecting the heat 
capacity $C$ of the black hole, its mass $M$ and temperature $T$ as given 
below:
\begin{equation*}
C=\frac{dM}{dT}.
\end{equation*}
From this relation, we derive the expression for the GUP-corrected heat 
capacity of the black hole as
\begin{equation}
C_{GUP}=-\frac{\pi  \beta \left(4 \alpha  (1-e)+r_H \left(\sqrt{\frac{(e-1)^2 \left((4 \alpha +r_H)^2-64 \beta \right)}{r_H^2}}-e+1\right)\right)\left((\alpha +4\, r_H)^2-4 \beta \right)}{8\, r_H \sqrt{\frac{(e-1)^2 \left((4 \alpha +r_H)^2-64 \beta \right)}{r_H^2}}\left((g-1) (\alpha +4\, r_H)^2+4 \beta \right)},
\label{eq23}
\end{equation}
where we have introduced a term $g$ defined as $g=\sqrt{1-\frac{4\beta}{(4r_H+\alpha)^2}}$ and considered $\omega=-\frac{1}{3}$. Fig.~\ref{fig7} shows the 
variation of heat capacity function \eqref{eq23} with the horizon radius for
different values of $e$, $\alpha$ and $\beta$ considering 
$\omega=-\frac{1}{3}$ and $M=1$. It is seen that the heat capacity is 
independent of the quintessence field, but depends heavily on the GUP-parameters
$\alpha$ and $\beta$. This dependence is more pronounced on the parameter 
$\beta$, especially for small horizon radii black holes. In almost all the 
cases the heat capacity is significantly different from the case of the 
Schwarzschild black hole. The heat capacity is negative throughout, which is 
very large for the small horizon radii black holes as well as for the cases 
of large radii ones. Thus GUP-corrected black holes should lose more energy 
in the form of radiation than that of the Schwarzschild black hole, in 
particular, by small and large horizon radii black holes.  

As already stated, when a black hole is not exchanging any heat with its 
surrounding, then we call this stable state as the remnant of the black 
hole. In this case, the heat capacity becomes zero. From the above expression, 
it can be shown that for the existence of remnant, the expression of the 
horizon radius of the remnant comes out as
\begin{equation}
r_{rem}=\frac{1}{4} \left(2 \sqrt{\beta }-\alpha \right).
\label{eq24}
\end{equation}
Thus, the remnant horizon radius depends on the deformation parameters $\alpha$ 
and $\beta$, and is independent of the behaviour of the surrounding field. It 
is interesting to note that in Ref.~\cite{bcl}, this dependency was 
established with one parameter only. The expression for the remnant 
temperature is calculated as
\begin{equation}
T_{rem}= \frac{3 e \omega  \left(\frac{\sqrt{\beta }}{2}-\frac{\alpha }{4}\right)^{-(3 \omega +1)}+1}{\pi  \sqrt{\beta }}.
\label{eq25}
\end{equation}
For instance, the remnant temperature for a particular combination of 
$\alpha=0.05$, $\beta=0.05$, $e=0.05$ and $\omega=-\frac{1}{3}$ comes out to 
be $1.352$, which is above the upper limit of $T_{GUP} \le 1.210$ for this 
case as obtained from the equation \eqref{eq21}. This upper limit of $T_{GUP}$
is calculated from the condition that 
$$\sqrt{1-\frac{4\beta}{(4r_H+\alpha)^2}}\;\;\ge 0,$$ which gives minimum 
allowed horizon radius for this case as $\sim 0.1$. This implies that the 
GUP-corrected Schwarzschild black holes in our study can not reach the stage 
of the remnant, which is also clear from the heat capacity analysis above.

The entropy function of the black hole can be estimated from the thermodynamic 
relation,
\begin{equation}
S=\int\frac{dM}{T}
\label{eq26}
\end{equation}
which, with the help of equations \eqref{eq8} and \eqref{eq21}, can be 
expressed for the GUP-corrected black hole as
\begin{equation}
S_{GUP}= \frac{\pi  M^2 \left((a+1) (\alpha +4 r_H)^2-4 \beta  \log ((a+1) (\alpha +4 r_H))\right)}{32 \left(M^2-4 \beta  (e-1)^2\right)}.
\label{eq27}
\end{equation}
\begin{figure}[h!]
\includegraphics[scale=0.28]{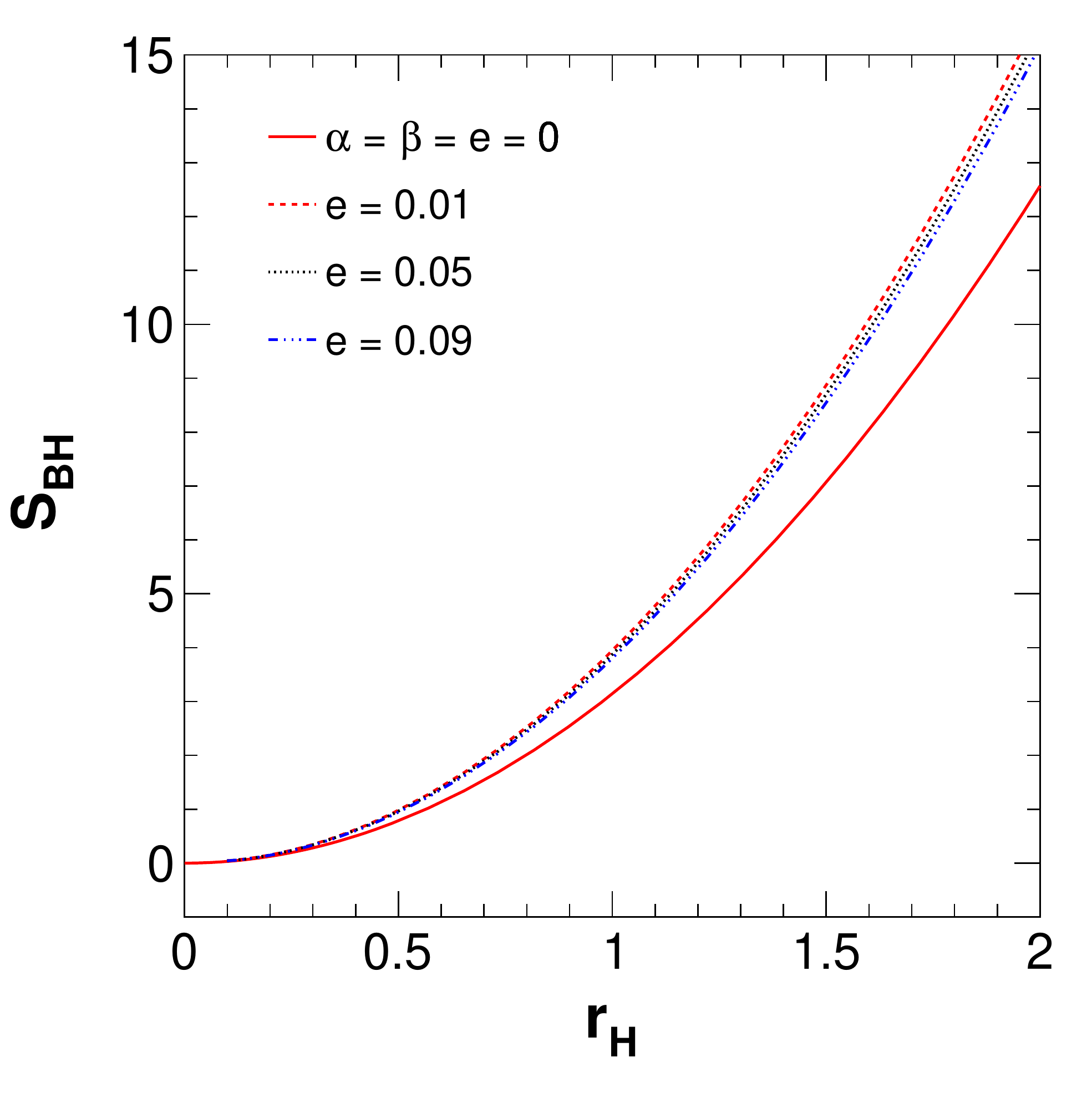}\hspace{0.3cm}
\includegraphics[scale=0.28]{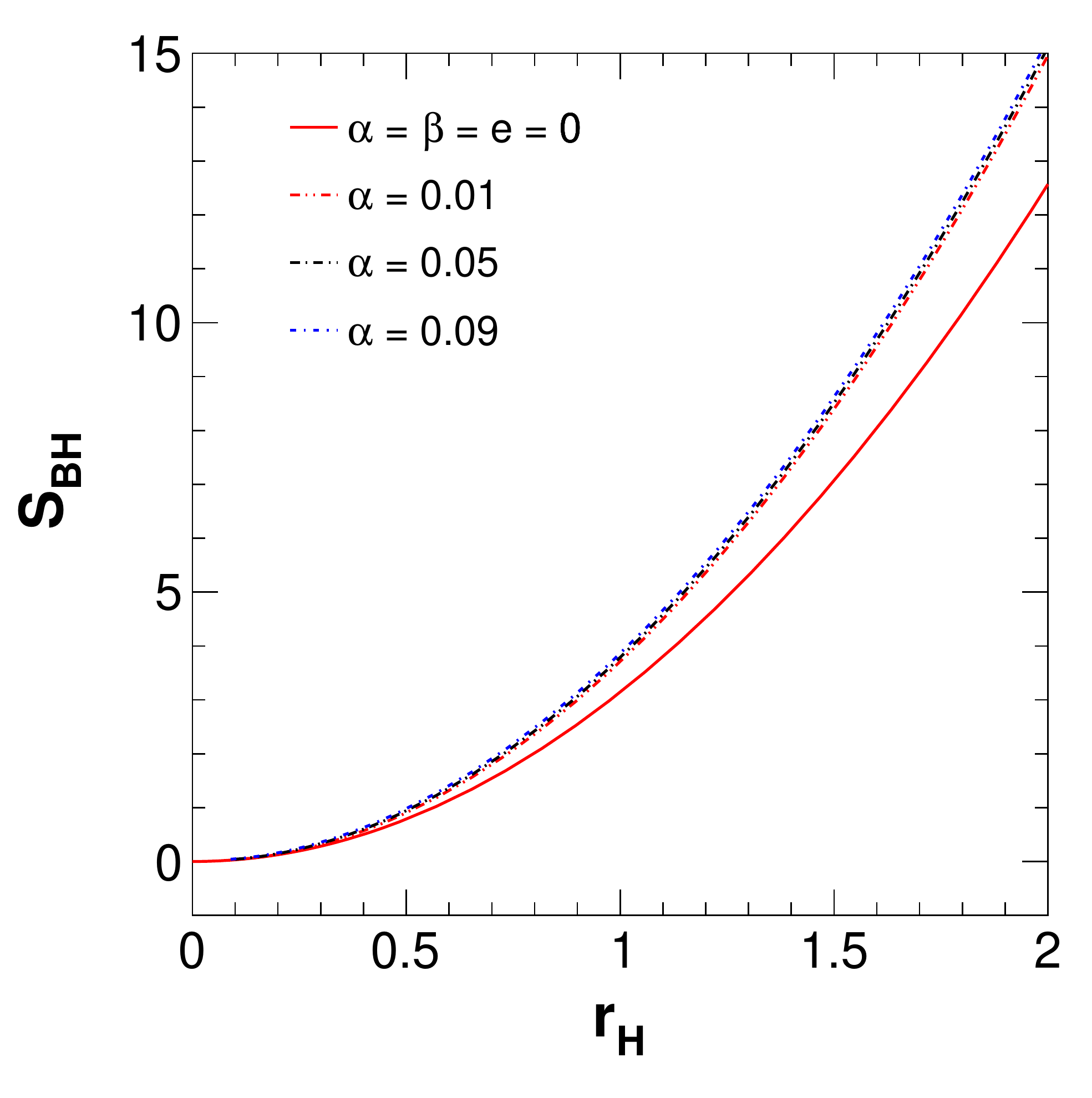}\hspace{0.3cm}
\includegraphics[scale=0.28]{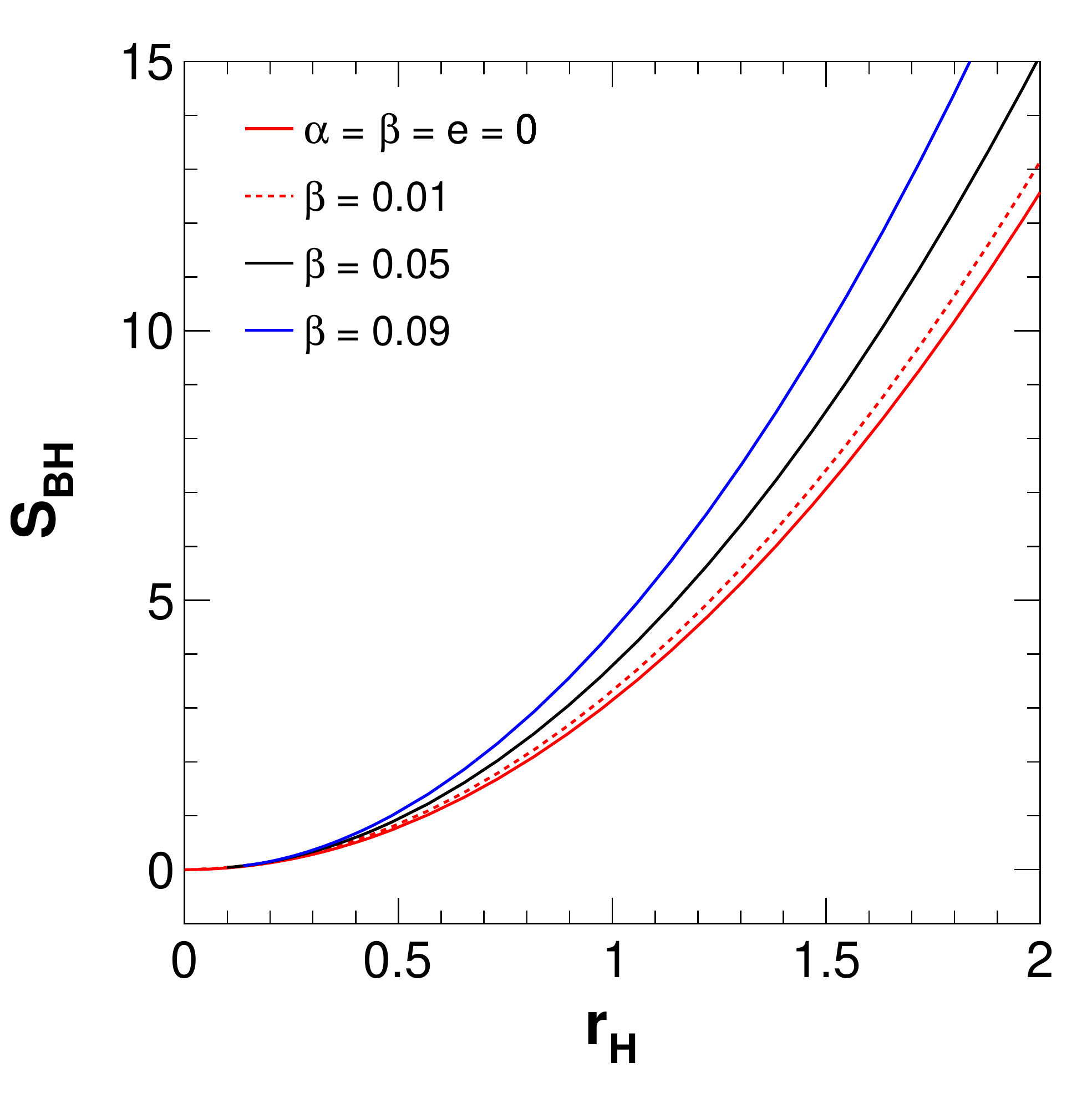}
\caption{Variation of entropy of GUP-corrected black holes with respect to
horizon radius for different values of $e$ with $\alpha = \beta = 0.05$ 
(left plot), for different values of $\alpha$ with $e=0.1$ and $\beta =0.05$
(middle plot), and for different values of $\beta$ with $e=0.1$ and 
$\alpha =0.05$ (right plot). In all three plots the solid red line
represents the entropy of Schwarzschild black holes.}
\label{fig7ad}
\end{figure}
Fig.~\ref{fig7ad} shows the variations of the entropy function \eqref{eq27}
with respect to the horizon radius of the GUP-corrected black holes for 
different values of the model's parameters together with the same for the 
Schwarzschild black holes. It is observed that the parameter $e$ and 
$\alpha$ have a very negligible impact on the entropy of the black holes. 
Whereas, the parameter $\beta$ has a significant impact on the entropy of 
black holes with sufficient horizon radii \cite{bcl}. Moreover, the entropy of
the GUP-corrected black holes is found to be higher than that of the 
Schwarzschild black holes for almost all cases and for sufficient horizon 
radius. This difference is substantial for a black hole with a larger horizon
radius depending on the value of model parameter $\beta$.   

\section{Conclusion} \label{conclusion}
The primary objective of this work is to study the effects of the 
deformation parameters introduced by the GUP on the QNMs of oscillation 
of the Schwarzschild black holes, together with a brief review of the 
thermodynamic properties of such GUP-corrected black holes surrounded by 
a quintessence dark energy field. It has been observed that both the 
deformation parameters as well as the quintessence parameter play an important 
role on the behaviour of the QNMs of the black holes. We employed two methods 
for obtaining the QNMs, namely the Mashhoon method and the 6th order WKB method. Further 
we derive a GUP-modified temperature expression of the black holes and show 
its dependence on the deformation parameters as well as on the quintessence
parameter. It is seen that there exist an upper bound on the temperature, 
which is impacted by the deformation parameters. Further, the heat capacity 
along with entropy have been evaluated for the GUP-corrected black holes and 
existence of black hole remnants has been studied. The existence of remnant 
radius and remnant temperature are certainly impacted by the deformation 
parameters. We observed that the GUP-corrected black holes can not reach the 
state of remnant. It is also seen that the quintessence field and the first
deformation parameter have no effective roles in the entropy of the black 
holes, which is dependent only on the second deformation parameter. It is quite remarkable that the 
introduction of small quantum corrections to the black hole metric can 
have notable influence on various properties of the black holes. This avenue 
has been investigated in the literature for many years but there are further 
scopes in this direction apart from the present study. It will be interesting 
to analyse the impact of these deformations on black hole properties, 
considering various Modified Theories of Gravity (MTGs). 

It is to be noted that once perturbed, a black hole responds by radiating 
GWs, which evolve in time. This evolution process is divided into three 
phases, an instant outburst of radiation, a longer period of damped 
oscillations (QNMs) and at very late times, a suppressed power-law 
tail \cite{konoplya_1}. Since we have been able to detect the first phase of GW 
only, it remains a challenge 
for the physicists and engineers to develop and improve the sensitivity of 
modern day detectors so that the second phase of GW can be detected. Steps in 
this direction have already been undertaken in the means of ambitious future 
projects like the LISA \cite{A8} and the Einstein Telescope \cite{A9}, which 
is believed to have far better sensitivity than the present detectors. The
prospect of detection of QNMs by LISA has been analysed in Ref.~\cite{gogoi4}. 
The detection of QNMs of the black hole can have many useful implications. It 
can be used to constrain the GUP parameters and as testing grounds for various 
MTGs as well. These upcoming advanced detectors of GWs can shed more light into 
this field and provide data for validating various models available at present 
times, which is one of the primary objectives in this field of study.

\section*{Acknowledgement} UDG is thankful to the
Inter-University Centre for Astronomy and Astrophysics (IUCAA), Pune, India
for the Visiting Associateship of the institute.

\appendix* \label{Appendix}
\section{Analytical form of expressions for QNM calculations by Mashhoon Method}

The analytical formula for the QNM frequency calculation by Mashhoon method is 
as follows:
$$\omega(l,n)=\omega_{R}+ i \omega_{I}=\Big(V_{0}-\frac{a^{2}}{4}\Big)^{\frac{1}{2}}+ia\Big(n+\frac{1}{2}\Big),$$ 
where the expreesions for the parameters $V_0$, $a$ and $x_{0}$ 
are found as

\begin{align*}
\boxed{
V_0=-\frac{4 A_1 A_3 (e-1)^3 l^3 (l+1)^3 M^4}{A_2^4},
}
\end{align*}
\begin{align*}
\boxed{
a=\sqrt{-\frac{A_7 A_9 \Big(2 \alpha  A_8 M^2+3 H \big(e+l^2+l-1\big)+4 \beta  (e-1)^2 M T+M^3 T\Big)}{A_5 A_6}},
}
\end{align*} 
\begin{equation*}
\begin{aligned}
x_0=\frac{1}{2 \Big(2 e l^2 M^2+2 e l M^2-2 l^2 M^2-2 l M^2\Big)}\Big[-24 \beta  e^3 M-24 \beta  e^2 l^2 M-24 \beta  e^2 l M-\\12 \alpha  e^2 M^2+72 \beta  e^2 M-\big[\big(24 \beta  e^3 M+24 \beta  e^2 l^2 M+24 \beta  e^2 l M+12 \alpha  e^2 M^2-72 \beta  e^2 M+\\12 \alpha  e l^2 M^2-48 \beta  e l^2 M+12 \alpha  e l M^2-48 \beta  e l M+6 e M^3-24 \alpha  e M^2+72 \beta  e M+6 l^2 M^3-\\12 \alpha  l^2 M^2+24 \beta  l^2 M+6 l M^3-12 \alpha  l M^2+24 \beta  l M-6 M^3+12 \alpha  M^2-24 \beta  M\big)^2-4 \big(2 e l^2 M^2+\\2 e l M^2-2 l^2 M^2-2 l M^2\big)  \big(256 \beta ^2+ 256 \beta ^2 e^4- 1024 \beta ^2 e^3+256 \alpha  \beta  e^3 M+1536 \beta ^2 e^2\\+64 \alpha ^2 e^2 M^2+128 \beta  e^2 M^2-768 \alpha  \beta  e^2 M-1024 \beta ^2 e+64 \alpha  e M^3-128 \alpha ^2 e M^2-256 \beta  e M^2+ \\768 \alpha  \beta  e M+16 M^4-64 \alpha  M^3+64 \alpha ^2 M^2+128 \beta  M^2-256 \alpha  \beta  M\big)\big]^{\frac{1}{2}}-12 \alpha  e l^2 M^2+48 \beta  e l^2 M\\-12 \alpha  e l M^2+48 \beta  e l M-6 e M^3+24 \alpha  e M^2-72 \beta  e M-6 l^2 M^3+12 \alpha  l^2 M^2-24 \beta  l^2 M-6 l M^3+\\12 \alpha  l M^2-24 \beta  l M+6 M^3-12 \alpha  M^2+24 \beta  M\Big].
\end{aligned}
\end{equation*}
Here, we make use of various definitions as
\begin{equation*}
\begin{aligned}
H = \Big[(9 + 9 e^2 + 14 l + 23 l^2 + 18 l^3 + 9 l^4 - 
    2 e (9 + 7 l + 7 l^2)) M^2 (M^2 +\\ 2 (-1 + e) M \alpha + 
    4 (-1 + e)^2 \beta)^2\Big]^{\frac{1}{2}},
 \end{aligned}   
 \end{equation*}
 $$T=9 e^2-2 e \left(7 l^2+7 l+9\right)+9 l^4+18 l^3+23 l^2+14 l+9,$$
    
 $$ A_1 = (L M^3 + 
  2 (3 + 3 e^2 + l + l^2 - e (6 + l + l^2)) M^2 \alpha + 
  4 (-1 + e)^2 L M \beta + H),$$
$$  A_2 = 6 \alpha  M^2 \Big(e^2+e \big(l^2+l-2\big)-l^2-l+1\Big)+3 M^3 C+12 \beta (e-1)^2 M C+H,$$
\begin{equation*}
\begin{aligned}
  A_3 = {}& -2 \alpha  M^2 \Big(e^2-e \big(3 l^2+3 l+2\big)+3 l^2+3 l+1\Big)+M^3 J+4 \beta  (e-1)^2 M \\& J+H,
\end{aligned}
\end{equation*}
\begin{equation*}
\begin{aligned}
A_4=4 (e-1)^3 l^3 (l+1)^3 M^4 \Big(-2 \alpha  M^2 \big(e^2-e (3 l^2+3 l+2)+3 l^2+3 l+1\big)+4 \beta  (e-1)^2 J M+H+\\J M^3\Big) \Big(2 \alpha  M^2 \big(3 e^2-e (l^2+l+6)+l^2+l+3\big)+4 \beta  (e-1)^2 M L+H+L M^3\Big),
\end{aligned}
\end{equation*} 
$$A_5=\frac{2 (-A_4)}{\Big(6 \alpha  M^2 \big(e^2+e (l^2+l-2)-l^2-l+1\big)+12 \beta  (e-1)^2 M P+H+3 M^3 P\Big)^4},$$
$$A_6=\Big(6 \alpha  M^2 \big(e^2+e (l^2+l-2)-l^2-l+1\big)+3 M^3 P+12 \beta  (e-1)^2 M P+H\Big)^8,$$
\begin{equation*}
\begin{aligned}
A_7=\Big(6 \alpha  M^2 \big(e^2+e (l^2+l-2)-l^2-l+1\big)+M^3 \big(e (4 l^2+4 l+3)-l^2-l-3\big)+1\\ 2 \beta  (e-1)^2 M (e+l^2+l-1)+H\Big)^2,
\end{aligned}
\end{equation*}
$$A_8=9 e^3-e^2 \big(14 l^2+14 l+27\big)+e \big(9 l^4+18 l^3+37 l^2+28 l+27\big)-9 l^4-18 l^3-23 l^2-14 l-9,$$
$$A_9=64 (e-1)^5 l^5 (l+1)^5 M^9 \big(4 \beta  (e-1)^2+2 \alpha  (e-1) M+M^2\big),$$
$$L = 3 c-l^2-l-3,$$
$$P=c+l^2+l-1,$$
$$J=-c+3 l^2+3 l+1,$$
$$C=c+l^2+l-1.$$



\bibliographystyle{apsrev}

\end{document}